\documentclass[prx,twocolumn,unsortedaddress,superscriptaddress]{revtex4-2}
\usepackage[english]{babel}
\addto\captionsenglish{}
\usepackage{graphicx}

\usepackage{epsfig,color}
\usepackage{amsmath,amssymb,eufrak,calc}
\usepackage{amsthm}
\usepackage{enumerate}
\usepackage{hyperref}
\usepackage{amsxtra,amsfonts,bm,txfonts}
\usepackage{blkarray}
\usepackage{adjustbox}
\usepackage{braket}
\usepackage{adjustbox}
\usepackage{xcolor}

\newcommand{\fulleqref}[1]{Eq. \eqref{#1}} % equation reference
\newcommand{\figref}[1]{Fig. \ref{#1}} % figure reference
\newcommand{\fullcite}[1]{Ref. \cite{#1}} % citation reference
\newcommand{\paren}[1]{\left( {#1} \right)} % ()
\newcommand{\parsq}[1]{\left[ {#1} \right]} % []
\newcommand{\parcr}[1]{\left\{ {#1} \right\}} % {}
\newcommand{\Qbraket}[2]{\left\langle {#1} \middle| {#2} \right\rangle}
\newcommand{\Qmatrixelement}[3]{\left\langle {#1} \middle| {#2} \middle| {#3} \right\rangle}
\usepackage{siunitx}

\begin{document}

\title{Resource-Efficient Quantum Simulation of Lattice Gauge Theories in Arbitrary Dimensions: Solving for Gauss' Law and Fermion Elimination}

\date{\today}

\author{Guy Pardo}
\address{Racah Institute of Physics, The Hebrew University of Jerusalem, Jerusalem 91904, Givat Ram, Israel.}

\author{Tomer Greenberg}
\address{Racah Institute of Physics, The Hebrew University of Jerusalem, Jerusalem 91904, Givat Ram, Israel.}

\author{Aryeh Fortinsky}
\address{Racah Institute of Physics, The Hebrew University of Jerusalem, Jerusalem 91904, Givat Ram, Israel.}

\author{Nadav Katz}
\address{Racah Institute of Physics, The Hebrew University of Jerusalem, Jerusalem 91904, Givat Ram, Israel.}

\author{Erez Zohar}
\address{Racah Institute of Physics, The Hebrew University of Jerusalem, Jerusalem 91904, Givat Ram, Israel.}

\begin{abstract}
Quantum simulation of Lattice Gauge Theories has been proposed and used as a method to overcome theoretical difficulties in dealing with the non-perturbative nature of such models. In this work we focus on two important bottlenecks that make developing such simulators hard: one is the difficulty of simulating fermionic degrees of freedom, and the other is the redundancy of the Hilbert space, which leads to a waste of experimental resources and the need to impose and monitor the local symmetry constraints of gauge theories. This has  previously been tackled in one dimensional settings, using non-local methods. Here we show an alternative procedure for dealing with these problems, which removes the matter and the Hilbert space redundancy, and is valid for higher space dimensions. We demonstrate it for a $\mathbb{Z}_2$ lattice gauge theory and implement it experimentally via the IBMQ cloud quantum computing platform.
\end{abstract}

\maketitle
%%%%%%%%%%%%%%%%%%%%%%%%%%%%%%%% INTRODUCTION %%%%%%%%%%%%%%%%%%%%%%%%%%%%%%%%%%%
\section{Introduction}
Gauge theories, describing the fundamental interactions among the constituents of matter, pose a serious challenge. Many are non-perturbative, at least for some energy scales; e.g., Quantum Chromodynamics (QCD), the theory of the strong nuclear force, is asymptotically free in high energies \cite{bjorken_asymptotic_1969,gross_ultraviolet_1973}, but non-perturbative at low energies. Perturbative techniques fail to describe this regime which exhibits very important physical phenomena, such as  quark confinement \cite{wilson_confinement_1974} which is responsible for the hadronic structure. This strongly coupled physics has been successfully addressed (see e.g. \cite{aoki_flag_2020}) by applying Monte-Carlo methods to lattice gauge theories (LGTs) \cite{wilson_confinement_1974,kogut_hamiltonian_1975,kogut_introduction_1979} - lattice formulations of gauge theories. However, these methods cannot directly describe real-time dynamics (being based on Euclidean time) or the physics of fermions with  finite chemical potentials, due to the sign problem \cite{troyer_computational_2005}. Thus, quantum simulation \cite{feynman_simulating_1982}, where a hard-to-solve quantum problem is mapped to a highly controllable quantum device which can be studied experimentally, would be useful in this case.

Recently, different approaches for quantum simulation of LGTs have been introduced (see, e.g., the reviews \cite{wiese_towards_2014,zohar_quantum_2016,dalmonte_lattice_2016,banuls_review_2020,banuls_simulating_2020,klco_standard_2021,zohar_quantum_2022,aidelsburger_cold_2022}), and implemented experimentally (e.g.  \cite{martinez_real-time_2016,kokail_self_2019,schweizer_floquet_2019,mil_scalable_2020,yang_observation_2020,semeghini_probing_2021,zhou_thermalization_2021}). Despite the enormous amount of work in the field, quantum simulation of LGTs remains a challenging endeavor. In particular, the complicated formulation of gauge theories imposes serious requirements on the simulated physics which call for creative simulation techniques, especially in more than one spatial dimension \cite{zohar_quantum_2022}. The reasons for that are numerous.

First, the matter is usually fermionic, and the gauge field is not. This entails combining fermionic and non-fermionic ingredients in the simulator, which often leads to choosing ultra-cold atomic systems for simulations \cite{zohar_quantum_2016}. In a single space dimension ($d=1$) this can be overcome using the Jordan-Wigner map \cite{jordan_uber_1928} which replaces fermions by spins - but introduces non-locality. Second, LGTs are highly constrained: the local symmetry introduces conservation laws (\emph{Gauss' law}) on every site, giving rise to a redundancy in the Hilbert space, which requires the simulation of unnecessary degrees of freedom, wasting costly resources. Finally, in $d>1$, the LGT Hamiltonian introduces the nontrivial four-body \emph{ plaquette  interaction} \cite{kogut_hamiltonian_1975}, which is not possessed naturally by the common quantum devices.

The latter issue may be dealt with in several ways, and in particular using a Trotterized approach \cite{trotter_on_1959,suzuki_decomposition_1985}: instead of mapping the Hamiltonian of the simulated model to that of the simulator (which is known as \emph{analogue quantum simulation}), one approximates the time evolution operator $\exp{\paren{-iHt}}=\exp{\paren{-i\underset{i}{\sum}H_it}}$ by a sequence of short time unitaries: $\exp{\paren{-i \epsilon H_i }}$, for $\mathcal{N}=t/\epsilon$ large enough such that 

\begin{equation} \label{eqn:Trotter}
 e^{-iHt} \approx \left(\underset{i}{\prod}e^{-i \epsilon H_i }\right)^\mathcal{N}.   
\end{equation}
By implementing each $H_i$ individually,  complicated interactions may be composed out of two-body unitaries, possibly using auxiliary ingredients. This is useful, in particular (but not only), for the four-body plaquette interactions included in LGTs \cite{zohar_digital_2017,zohar_digital_2017-1,bender_digital_2018,lamm_general_2019,armon_photon_2021,gonzalez_hardware_2022}.

Here we address the first two issues by using a reformulation of lattice gauge theories which uses the local constraints to eliminate the fermionic matter  \cite{zohar_eliminating_2018,zohar_removing_2019}. Here we address the first two issues by using a reformulation of lattice gauge theories which uses the local constraints to eliminate the fermionic matter [33,34]. We construct a quantum simulation algorithm based on it, show specifically how to apply it to Z2 LGT to build a working quantum simulation, and demonstrate an experimental implementation of it. Our protocol yields a \emph{local} model that does not include matter, but is still equivalent to the original  LGT (with fermions); as a result, not only  do we not need to simulate any fermions directly, but there are also no local constraints to impose and maintain and no redundancy in the Hilbert space. We thus obtain a much simpler simulation scheme that is valid also for $d>1$.

For the sake of completeness, we would like to mention that other methods, using different types of tools, are also used in order to address the three issues mentioned above. These include, for example, the loop-string-hadron formalism, which formulates lattice gauge theories in terms of explicitly gauge invariant degrees of freedom, allowing one to remove the constraints \cite{raychowdhury_loop_2020,kadam_loop_2022} (see \cite{davoudi_search_2021} for a comparative study of this and other methods for the quantum simulation of $SU(2)$ models in a single space dimension). Another approach is the use of dual formulations, which allow, at least in the Abelian case, to switch to magnetic degrees of freedom, which are free of the Gauss law constraints and have no plaquette interactions, but do not directly address the issue of fermionic matter \cite{drell_quantum_1979,kaplan_gauss_2018,bender_gauge_2020,haase_resource_2021,paulson_simulating_2021,bauer_efficient_2021}.

We would also like to mention, that after the completion of the first version of this article, we became aware of a parallel work on simulating $\mathbb{Z}_2$ lattice gauge theories using the same methods \cite{irmejs_quantum_2022}. The analysis performed in the two works may be seen as complementary.

The article is organized as follows. We begin by reviewing the basics of Hamiltonian LGTs (section \ref{LGT_review}), focusing on $\mathbb{Z}_2$ as the simplest case, which already shows the relevant features (redundancy of the Hilbert space and fermionic matter). In section \ref{method_0}, we  review the conventional non-local, $1+1d$ techniques to deal with these issues \cite{hamer_lattice_1979,martinez_real-time_2016,banuls_efficient_2017,sala_variational_2018,atas_su_2021} and apply them to $\mathbb{Z}_2$. The main step involves a unitary transformation that we introduce and denote as $\mathcal{U}^{(0)}$. Section \ref{1d_theory} introduces our local procedure and applies it for $\mathbb{Z}_2$ in $d=1$. The procedure involves two unitary steps that we introduce and denote as $\mathcal{U}^{(1)}$ and $\mathcal{U}^{(2)}$. We proceed to presenting a few experimental demonstrations (section \ref{1d_experiment}) of the method implemented on IBMQ devices. Next, we generalize our procedure to $d=2$ (section \ref{sec:2d_theory}), and present an experimental implementation of a quasi-two-dimensional system (section \ref{2d_experiment}) which is the simplest system for which the standard method of $\mathcal{U}^{(0)}$ fails. Finally, in section \ref{discussion_and_summary} we use numerical simulations to estimate the near-term experimental feasibility of using our method for more advanced applications than those presented in section \ref{1d_experiment}.

%%%%%%%%%%%%%%%%%%%%%%%%%%%%%%%% BACKGROUD %%%%%%%%%%%%%%%%%%%%%%%%%%%%%%%%%%%
\section{Background}\label{background}
\subsection{Hamiltonian lattice gauge theories} \label{LGT_review}
\begin{figure} 
	\includegraphics[width=0.8\columnwidth]{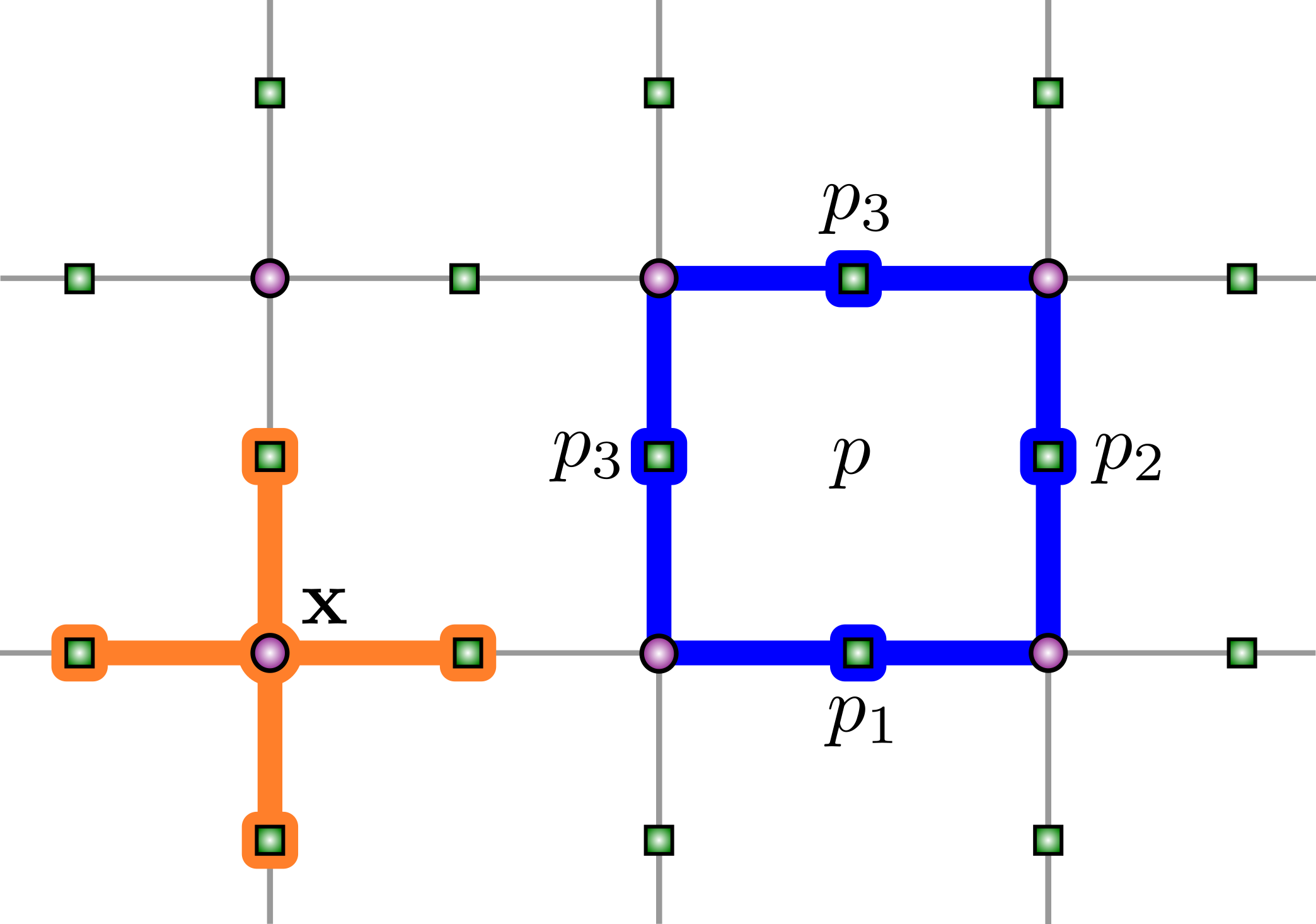}
	\caption{The LGT configuration space of: matter (purple circles)  on the  sites,  gauge fields (green squares)  on the links. The highlighted degrees of freedom on the left are those on which gauge trasformations $\Theta\left(\mathbf{x}\right)$ act; the highlighted plaquette on the right sets the $H_\text{B}$ convention of the text (\fulleqref{eqn:H_B_def}).}
	\label{Fig1}
\end{figure}

Hamiltonian LGTs \cite{kogut_hamiltonian_1975} are defined on $d$ dimensional spatial lattices $\mathbb{Z}^d$ (square/cubic by default, though formulations in other geometries exist). LGTs include two types of fields: the matter, mostly (but not necessarily) fermionic, associated with the lattice sites and described by the fermionic Fock space $\mathcal{H}_{\text{m}}$, and the gauge field, associated with the lattice links and described by the Hilbert space $\mathcal{H}_{\text{g}}$ (see Fig. \ref{Fig1}). 
The gauge group $G$ is a compact Lie or a finite group that generates \emph{gauge transformations}: local unitaries  $\Theta_g\left(\mathbf{x}\right)$ under which the physically relevant states and operators are invariant. These are parametrized by group elements  $g \in G$,  and are associated with the  sites $\mathbf{x} \in \mathbb{Z}^d$. Each $\Theta_g\left(\mathbf{x}\right)$ acts locally on $\mathbf{x}$ and the links $\ell \ni \mathbf{x}$ around it (starting or ending at $\mathbf{x}$, see Fig. \ref{Fig1}), transforming only those degrees of freedom in a way that is parametrized by $g$. A gauge invariant operator $O$ satisfies
\begin{equation} \label{eqn:gauge_invariance}
 \Theta_g\left(\mathbf{x}\right) O \Theta^{\dagger}_g\left(\mathbf{x}\right) = O, \hspace{30pt} \forall g\in G,\mathbf{x}\in\mathbb{Z}^d;   
\end{equation}
and a gauge invariant state $\left|\psi\right\rangle$ is  invariant under all gauge transformations (up to a global phase if $G$ is abelian; in the non-Abelian case, gauge transformations can mix the elements of state multiplets  \cite{kasper_from_2020}). 

Let us focus on the case $G=\mathbb{Z}_2$. Each site can host a single fermion, annihilated by $\psi\left(\mathbf{x}\right)$. Each link hosts a two dimensional Hilbert space.  $\mathbb{Z}_2$ contains a single nontrivial group element; thus the possible gauge transformations are
\begin{equation}
\Theta\left(\mathbf{x}\right) = S\left(\mathbf{x}\right) e^{i\pi N\left(\mathbf{x}\right)},
\label{gtrans}
 \end{equation}
where 
$S\paren{\mathbf{x}} \equiv\left[\underset{\ell\ni \mathbf{x}}{\prod}Z\left(\ell\right) \right]$
is a product of Pauli $z$ operators $Z\left(\ell\right)$ acting on the links $\ell$ that are connected to the site $\mathbf{x}$,  and $N\left(\mathbf{x}\right) = \psi^{\dagger}\left(\mathbf{x}\right)\psi\left(\mathbf{x}\right)$ is the number operator at $\mathbf{x}$. 
The gauge invariant operators are $Z$ operators, products of $X$ operators along closed loops, and functions thereof; but also (functions of) the so-called \emph{mesonic strings}, which are operators of the form  
\begin{equation} \label{eqn:mesonic_string_general}
\psi^{\dagger}\left(\mathbf{x}\right) \underset{\ell \in \mathcal{C}}{\prod}X\left(\ell\right) \psi\left(\mathbf{y}\right), 
\end{equation} 
where $\mathcal{C}$ is any path connecting the sites $\mathbf{x}$,$\mathbf{y}$. Note that the mesonic strings include trivially the number operator $N\paren{\mathbf{x}}$.

A conventional Hamiltonian choice  \cite{kogut_hamiltonian_1975} takes the form
\begin{equation} \label{eqn:LGT_general_ham}
	H=H_{\text{E}}+H_{\text{B}}+H_{\text{GM}}+H_{\text{m}},
\end{equation}
where $H_{\text{E}}$, the \emph{electric energy}, is a sum of local gauge (electric) field terms on all the links $\ell$,  
the \emph{magnetic energy} $H_{\text{B}}$, is a four-body interaction of the  links around each plaquette (unit-square), and
$H_{\text{GM}}$ is the interaction with the matter that involves hopping of fermions to neighbouring sites, while changing the state of the field on the intermediate link.
$H_{\text{m}}$ is a the mass term, which we choose to be \emph{staggered} \cite{susskind_lattice_1977}, with generalizations to HEP-like LGTs in mind. 

While the Hamiltonian was originally formulated by Kogut and Susskind for continuous groups \cite{kogut_hamiltonian_1975}, following Wilson's Lagrangian formalism \cite{wilson_confinement_1974}, it is possible to extend it to finite groups, and in particular to 
$\mathbb{Z}_N$. Here, we follow the formulation of \cite{horn_hamiltonian_1979}, where the pure-gauge parts of the Hamiltonian are constructed such that they will give rise to the conventional $U(1)$ formulation in the large $N$ limit.
For the $\mathbb{Z}_2$ case the Hamiltonian terms can be written as:
\begin{equation} \label{eqn:H_E_def}
    H_{\text{E}}= -h\underset{\ell}{\sum}Z\left(\ell\right),
\end{equation}
\begin{equation} \label{eqn:H_B_def}
    H_{\text{B}}=  b\underset{p}{\sum}X_1\left(p\right)X_2\left(p\right)X_3\left(p\right)X_4\left(p\right),
\end{equation}
where the indices 1-4 label the four different links that form a given plaquette $p$  (see Fig. \ref{Fig1}),
\begin{equation} \label{eqn:H_GM_def}
    H_{\text{GM}}=-J\underset{\mathbf{x},i=1,...,d}{\sum}\psi^{\dagger}\left(\mathbf{x}\right)X\left(\mathbf{x},i\right)\psi\left(\mathbf{x}+\mathbf{e}_i\right)+\text{h.c.},
\end{equation}
where $\mathbf{e}_i$ is a lattice vector in direction $i$, and $X\left(\mathbf{x},i\right)$ acts on the link emanating from $\mathbf{x}$ in direction $i$; and 
\begin{equation} \label{eqn:H_m_def}
    H_{ \text{m}} = m \underset{\mathbf{x}}{\sum}\left(-1\right)^{x_1+...+x_d}N\left(\mathbf{x}\right),
\end{equation}
where the alternating sign is a result of staggering, such that for the odd sites the existence of a fermion ($N\paren{\mathbf{x}}=1$) can be interpreted as the vacuum (Dirac-sea) state, and the absence of a fermion ($N\paren{\mathbf{x}}=0$) can be interpreted as an anti-particle. The even sites follow the opposite and more intuitive convention where $N\paren{\mathbf{x}}=0$ represents the empty state and $N\paren{\mathbf{x}}=1$ represents a particle \cite{susskind_lattice_1977}. 

Gauge invariant states $\left|\psi\right\rangle$ satisfy the local \emph{Gauss' law } constraints, that for $\mathbb{Z}_2$ can be written as:
\begin{equation}
	\Theta\left(\mathbf{x}\right)\left|\psi\right\rangle = e^{i\pi q\left(\mathbf{x}\right)} \left|\psi\right\rangle, \quad \forall \mathbf{x} \in\mathbb{Z}^d.
	\label{Gauss}
\end{equation}
Since $H$ is gauge invariant, the eigenvalues $q\left(\mathbf{x}\right)=0,1$ are constants of motion, splitting the Hilbert space $\mathcal{H}$ into dynamically disconnected \emph{superselection sectors}, $\mathcal{H}\left(\left\{q\left(\mathbf{x}\right)\right\}\right)$. The physical Hilbert space satisfies
\begin{equation} \label{eqn:Hilber_space_decomposition}
	\mathcal{H}=\underset{\left\{q\left(\mathbf{x}\right)\right\}}{\bigotimes}\mathcal{H}\left(\left\{q\left(\mathbf{x}\right)\right\}\right) \subset \mathcal{H}_{\text{m}} \times \mathcal{H}_{\text{g}}.
\end{equation}
In a model with staggered mass as in \fulleqref{eqn:H_m_def}, the sector defined by $e^{i\pi q\paren{\mathbf{x}}} = \paren{-1}^{x_1 + ... +x_d}$ is often considered as the simplest sector in the sense that it includes the "Dirac-sea" state in which only odd sites are populated (no particles and no anti-particles). 

Since we are usually interested in a single sector, $\mathcal{H}_{\text{m}} \times \mathcal{H}_{\text{g}}$ is highly redundant, and implementing it would be very wasteful in resources. Implementations of $\mathbb{Z}_2$ LGTs with various settings have been discussed in \cite{zohar_digital_2017,schweizer_floquet_2019,barbiero_coupling_2019,cui_circuit_2020,gonzalez_robust_2020,homeier_Z2_2021,gustafson_toward_2021,semeghini_probing_2021,armon_photon_2021,mildenberger_probing_2022,samajdar_emergent_2022,homeier_quantum_2022,lumia_two_2022}. Here we deal with $\mathbb{Z}_2$ LGTs using other methods, specifically by removing the redundancy.

 \begin{figure} 
	\includegraphics[width=0.8\columnwidth]{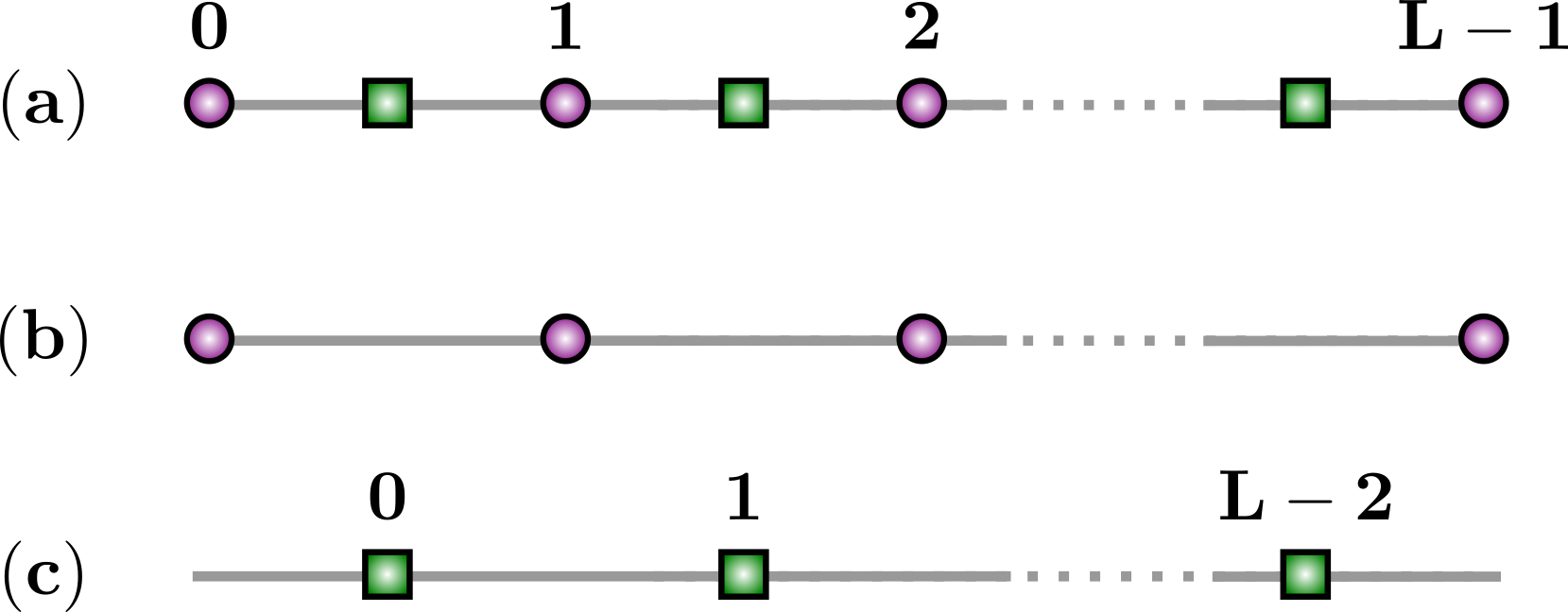}
	\caption{Comparison of configuration spaces of the $d=1$ model. (a) the original model with both the matter and gauge field degrees of freedom;  (b) When we eliminate the gauge field non-locally (section \ref{method_0}) we are left with a matter-only theory on the sites. (c) When the matter is eliminated locally (section \ref{1d_theory}), we are left only with the gauge fields on the links, which means that the model can be simulated by a chain of $L-1$ qubits with local interactions.}
	\label{Fig2}
\end{figure}

\subsection{Removing the redundancy in the standard approach: Eliminating the gauge fields}\label{method_0}
Removing the redundancy means solving Gauss' laws (\ref{Gauss}): 
\begin{equation} \label{eqn:Z2_gauss}
    \underset{\ell\ni \mathbf{x}}{\prod}Z\left(\ell\right) \left|\psi\right\rangle = e^{i\pi\left( N\left(\mathbf{x}\right)+q\left(\mathbf{x}\right)\right)}\left|\psi\right\rangle,
\end{equation}
for every lattice site $\mathbf{x}$.
In the standard approach, one solves it for the gauge fields: 
given $N\left(\mathbf{x}\right)$, we have to solve for $Z$ on each link. However, there are several $Z$ configurations satisfying \fulleqref{eqn:Z2_gauss}, unless $d=1$ (Fig. \ref{Fig2}(a)). In this case, we can label both the sites and the links by $n$,  and the constraints simplify to 
\begin{equation} \label{eqn:1d_gauss}
    Z_n Z_{n-1}\left|\psi\right\rangle = e^{i\pi \left(N_n+q_n\right)}\left|\psi\right\rangle
\end{equation}
which, for open boundary conditions ($0\leq n \leq L-1$, for an even number of sites $L$) is easily solved by the non-local expression:

\begin{equation}\label{eqn:gauss_solution_m0}
    Z_n\left|\psi\right\rangle = \exp{\left(i\pi \overset{n}{\underset{k=0}{\sum}}\left(N_k+q_k\right)\right)}\left|\psi\right\rangle.
\end{equation}

This motivates the unitary \emph{degauging} transformation:
\begin{equation} \label{eqn:U_0_def} % starting from  zero
    \mathcal{U}^{(0)} = \exp{\left(i\frac{\pi}{2}\underset{n}{\sum}\left(1-X_n\right)\overset{n}{\underset{k=0}{\sum}}\left(N_k+q_k\right)\right)}.
\end{equation}
We denote transformed states and operators as
\begin{align}
    \left|\psi\right\rangle&\xrightarrow{}\left|\psi^{(0)}\right\rangle=\mathcal{U}^{(0)}\left|\psi\right\rangle\\
    O&\xrightarrow{}O^{(0)}=\mathcal{U}^{(0)} O\mathcal{U}^{(0)\dagger}.
\end{align}
This transformation enforces the solution \eqref{eqn:gauss_solution_m0} in a given sector.

The transformed parts of the Hamiltonian are the interaction:
\begin{equation} 
    H^{(0)}_{\text{GM}} = \left(-J\overset{L-2}{\underset{n=0}{\sum}}\psi^{\dagger}_n\psi_{n+1} +\text{h.c.}\right),
\end{equation}
and the electric term, which becomes non-local:
\begin{equation}
    H^{(0)}_{\text{E}} = h\overset{L-2}{\underset{n=0}{\sum}}
\exp{\left(i\pi \overset{n}{\underset{k=0}{\sum}}\left(N_k+q_k\right)\right)}.  
\end{equation}
In $d=1$ there is no $H_\text{B}$ (no plaquettes), and the mass term is unaffected: $H^{\paren{0}}_\text{m} = H_\text{m}$.

\begin{figure} 
	\includegraphics[width=\columnwidth]{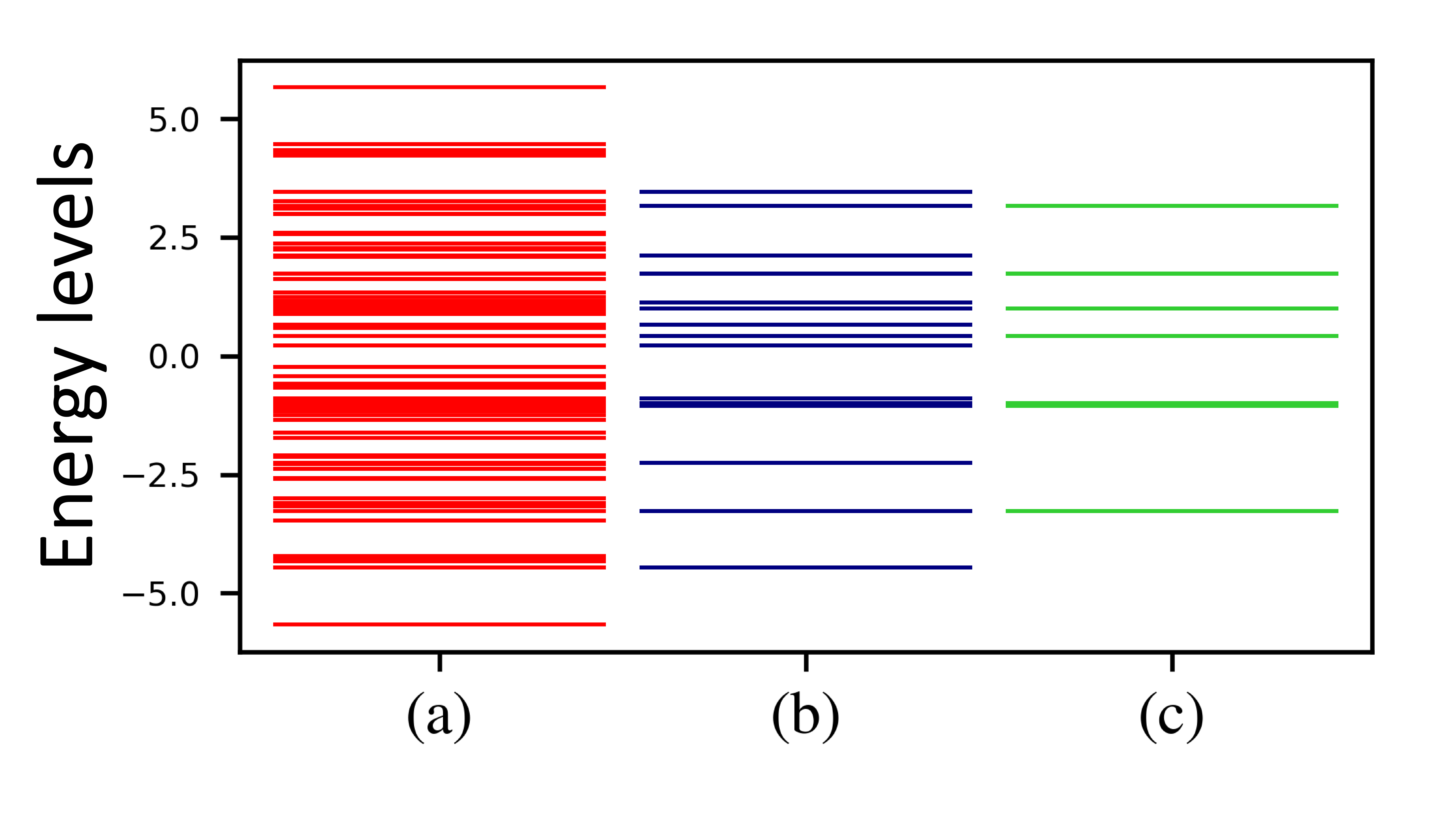}
	\caption{The spectra of (a) the full model, and after eliminating (b)  the gauge fields and (c)  the matter, for $h=J=m=1$. In the original formulation, the gauge fields and matter cannot be decoupled and the spectrum includes all the sectors. (b) and (c) correspond to transformed sectors of (a), with less degrees of freedom, and thus clearly contain less eigenstates.}
	\label{Fig3}
\end{figure}

Note that $\left[H^{(0)},Z_n\right]=0$ 
and $Z_n\left|\psi^{(0)}\right\rangle =\left|\psi^{(0)}\right\rangle$, $\forall n$.
Thus the redundancy is completely removed, the gauge fields are in a product state with the matter, and only the latter has to be treated in any quantum simulation scheme. The explicit presence of fermions is still a potential problem, restricting the choice of the simulating platform,  but in $d=1$ we can use the Jordan-Wigner transform \cite{jordan_uber_1928} to represent the fermions with qubits:
\begin{equation} \label{eqn:Jordan_Wigner}
\psi_n = \left[\overset{n-1}{\underset{k=0}{\prod}}\sigma_z^k\right]\sigma
^-_n,    
\end{equation}
which results in an $L$ qubit system, residing on the sites (Fig. \ref{Fig2}(b)). The Hilbert space dimension has decreased \emph{exponentially} from $2^{2L-1}$ to $2^L$, with no local constraints  left. This reduction is demonstrated in Figs. \ref{Fig3}(a,b), by comparing the spectra of $H$ and $H^{(0)}$. As $H^{(0)}$ is in a specific sector it contains less levels, but it describes the same physics as $H$ in the chosen sector.

For simplicity, we will focus from now on the sector where $e^{i\pi q_n} = \left(-1\right)^n$, in which the Hamiltonian is (up to a constant):
\begin{equation}
\begin{split}
        H^{(0)} &=\overset{L-2}{\underset{n=0}{\sum}}\left[  -h\left(-1\right)^{n\left(n+3\right)/2}
		\prod_{k=0}^{n}{\sigma^z_k}
	  + \left(J\sigma^+_n\sigma^-_{n+1} +\text{h.c.}\right)\right] \\
 &+ \frac{m}{2}\overset{L-1}{\underset{n=0}{\sum}} \left(-1\right)^{n}\sigma^z_n. \label{eqn:H0}
\end{split}
\end{equation}

The dynamics of $H^{(0)}_{\text{m}}$ can be simulated using local qubit rotations, and that of $H^{(0)}_{\text{GM}}$ by simple two-qubit gates on neighbouring sites. $H^{(0)}_{\text{E}}$, however, involves highly non-local many-body interactions, and thus it is more complicated and requires Trotterization, either to a strictly digital simulation or an analogue-digital one. In the latter, the evolution with respect to  $H^{(0)}_{\text{m}}$ and $H^{(0)}_{\text{GM}}$ can be implemented using analogue techniques (which is possible on some platforms). In any case, the local parts can be simulated with an $O(1)$ run-time.

To implement $e^{-i \epsilon H_{\text{E}}^{(0)}}$, we use two types of unitaries: (i) single qubit rotations,
$V_n = \exp\left(-i\epsilon h \left(-1\right)^{n\left(n+3\right)/2}\sigma^z_n\right)$; (ii)
CNOT gates between neighbouring qubits,
	$U_n = \left|\uparrow\right\rangle\left\langle\uparrow\right|_n 
	+ \left|\downarrow\right\rangle\left\langle\downarrow\right|_n \otimes \sigma^x_{n+1}$. Since 	$U_n \sigma^z_{n+1} U_n = \sigma^z_{n}\sigma^z_{n+1}$,
we get that 
\begin{equation}
e^{-i \epsilon H_{\text{E}}^{(0)}} = U_0 U_1 \cdots U_{L-3} V_{L-2} U_{L-3} V_{L-3}  U_{L-3} \cdots
U_1 V_1 U_0 V_0.
\end{equation}
Due to the non-locality, the length of this operation scales linearly in the system size $L$, and we conclude that quantum simulation of the dynamics using this method has an  $O(L)$ run-time per Trotter step.

While we demonstrated it for $\mathbb{Z}_2$, this way of integrating the gauge field out is valid for arbitrary gauge groups in $d=1$ \cite{hamer_lattice_1979,banuls_efficient_2017,sala_variational_2018,zohar_removing_2019}, and can be used for quantum simulation \cite{martinez_real-time_2016,atas_su_2021}. This method's drawbacks are that it is restricted to $d=1$, introduces non-locality and its Trotter step run-time depends on $L$. non-locality can arise in different ways, for example for Lie groups - see, e.g. \cite{martinez_real-time_2016}, where  the transformed electric Hamiltonian includes only two-body terms, but arbitrarily far; in that case, a successful experimental realization was possible using the long-range interactions of the simulating platform used (trapped ions).

%%%%%%%%%%%%%%%%%%%%%%%%%%%%%%% 1D THEORY %%%%%%%%%%%%%%%%%%%%%%%%%%%%%%%%%%
\section{Eliminating the matter}\label{1d_theory}
In our approach we solve the constraints as equations for the matter. This significantly simplifies the problem, since in this view these equations are explicitly solved: In the $\mathbb{Z}_2$ case, knowing the $Z$ configuration gives rise immediately to a unique  and local solution for $N\left(\mathbf{x}\right)$, and similar results are valid for other groups as well. Such a solution is quite straightforward for bosonic, Higgs-like matter (unitary gauge fixing) \cite{fradkin_phase_1979}. When we deal with fermions, things have to be done rather more carefully, but it is possible nevertheless. There are two steps to our procedure: first (section \ref{method1}) we transform the fermions into hard-core bosons, and then (section \ref{method2}) we solve Gauss' law for the matter and remove the redundancy, eliminating altogether the need to simulate the matter. In section \ref{trotterization_etc} we provide details on how to implement the quantum simulation (transformed) Hamiltonian using fully digital or hybrid techniques, as well as how to measure the gauge invariant observables.

\subsection{From fermions to hard-core bosons} \label{method1}
Following the procedure of \fullcite{zohar_eliminating_2018} we can replace the fermionic matter of any LGT whose gauge group contains $\mathbb{Z}_2$ as a normal subgroup by hard-core bosonic matter. After applying a unitary procedure $\mathcal{U}^{\text{(1)}}$ which preserves the physics of the original states $\left|\psi\right\rangle$, as given in \cite{zohar_eliminating_2018}, one ends up with equivalent states,
\begin{equation}
    \left|\psi^{(1)}\right\rangle = \mathcal{U}^{\text{(1)}}\left|\psi\right\rangle,
\end{equation}
 of a model in which each fermionic mode $\psi$ is replaced by a spin, or a hard-core boson; thanks to the local constraints (Gauss' law), the gauge field absorbs the statistics, leaving us only with non-fermionic matter fields and modifying slightly the way that the gauge fields appear in the Hamiltonian, to account for the statistics. Since this is done via local unitaries which exploit the local constraints, we are left with a local theory nevertheless \cite{zohar_eliminating_2018}. 

While, as shown in \cite{zohar_eliminating_2018}, the method is valid for any space dimension we will focus here on $d=1$ for the clarity of the presentation, and for comparison with the previous method (we treat the $d=2$ case in section \ref{sec:2d_theory}). On the other hand, we will switch from now on to periodic boundary conditions, which are simpler to deal with and, unlike when the gauge field is removed, are valid here. For $\mathbb{Z}_2$, when performing the procedure of \cite{zohar_eliminating_2018} and replacing the fermions by hard-core bosons, we get

\begin{equation} \label{eqn:H_1_1d}
    H^{(1)} =\underset{n}{\sum}\left[  -h Z_n
		 + \left(iJ Z_{n-1}\sigma^+_nX_n\sigma^-_{n+1} +\text{h.c.}\right) +\frac{m}{2}\left(-1\right)^{n}\sigma^z_n\right],
\end{equation}
where $\sigma^{\pm}_n$ are spin raising and lowering  operators for the hard-core bosonic mode at the site $n$.

Transforming the original Gauss' law (given by \fulleqref{eqn:1d_gauss}) with the substitution $N_n \rightarrow \sigma^+_n\sigma^-_n$, we obtain its hard-core bosonic form,
\begin{equation} \label{eqn:gauss_law_method1}
	-\varepsilon_n S_n\sigma^z_n\left|\psi^{(1)}\right\rangle = \left|\psi^{(1)}\right\rangle,
\end{equation}
where $S_n = Z_{n-1}Z_{n}$, and $\varepsilon_n \equiv e^{i \pi q_n}=\pm 1$ such that each choice of signs $\parcr{\varepsilon_n}_{n=0}^{L-1}$ defines a superselection sector.
 
The Hamiltonian $H^{\paren{1}}$ is free of fermions, but is subject to local constraints. The redundancy problem is unsolved, but one can still formulate a Trotterized quantum simulation of $H^{(1)}$, using $2L$ qubits ($2L-1$ for  open  boundaries). The usual set of gates and tools allows us to formulate it quite easily, but we still need to use a redundant Hilbert space and make sure that the constraints are satisfied, either by monitoring them directly \cite{stryker_oracles_2019,halimeh_stabilizing_2022,mildenberger_probing_2022} or making sure that each Trotter step is gauge invariant \cite{zohar_digital_2017,armon_photon_2021}.
  In the recent work \cite{mildenberger_probing_2022}, the $d=1$ $\mathbb{Z}_2$ was simulated without integrating out any degree of freedom (the fermions were taken care of by a Jordan-Wigner transform instead, and thus extending it to higher dimensions might be very challenging and non-local. Such ideas have been studied recently, e.g. in \cite{li_higher_2022} and references therein). 

\subsection{Solving Gauss' law for the matter} \label{method2}
Here, we shall proceed to a complete elimination of the matter, in a procedure similar to that of \fullcite{zohar_removing_2019}, using the fact that Gauss' law provides us with a one-to-one map between the values of $S_n$ and those of $\sigma_n^z$. An alternative approach for eliminating the matter, valid for $\mathbb{Z}_2$ only but giving rise to similar results, was studied in \cite{borla_confined_2020,borla_quantum_2020}; similar methods for eliminating matter in $d=1$ abelian systems were discussed in \cite{surace_lattice_2020,lerose_quasilocalized_2020,surace_scattering_2021}. Our procedure is valid for other gauge groups (including non-Abelian ones) and higher dimensions as well. Originally, it was given for $U(N)$ groups, and we shall now adapt it to the case of $\mathbb{Z}_2$, which was not explicitly included in \cite{zohar_removing_2019}.

First, we define  projectors onto $S_n$ eigenstates
\begin{equation}
    P_n^{\pm} = \frac{1}{2} \left(1\pm\paren{-\varepsilon_n} S_n\right),
\end{equation}
where $\varepsilon_n =\pm 1$, depending on the static charge sector.
and rewrite Gauss' law as
\begin{equation} \label{eqn:Gauss_with_projectors}
\left(P_n^{+}-P_n^{-}\right)\left|\psi^{(1)}\right\rangle = \sigma^z_n\left|\psi^{(1)}\right\rangle,     
\end{equation}
valid for any choice of signs $\parcr{\varepsilon_n}$. Define, on each site $n$, a controlled local unitary which decouples the matter: Suppose we want all the matter spins to point down. If $S_n = \varepsilon_n$, we do nothing, and if  $S_n = -\varepsilon_n$ we invert it (refer to \fulleqref{eqn:gauss_law_method1}). The controlled unitary that performs this operation is:
\begin{equation} \label{eqn:method2_unitary}
    \mathcal{U}_n = P^{+}_n \sigma_n^x + P^-_n,
\end{equation}
and since $\left[\mathcal{U}_n,\mathcal{U}_m\right]=0$, we can safely define 
\begin{equation}
    \mathcal{U}^{(2)} = \underset{n}{\prod}\mathcal{U}_n.
\end{equation}
This is the second unitary step in our procedure, and we denote transformed states and operators as 
\begin{equation}
    \left|\psi^{(2)}\right\rangle= \mathcal{U}^{(2)}\left|\psi^{(1)}\right\rangle   
\end{equation}
and $O^{(2)}= \mathcal{U}^{(2)}O^{(1)}\mathcal{U}^{(2)\dagger}$.
Note that the operators $\mathcal{U}_n^{(2)}$ depend on the projection operators $P^{\pm}_n$ which depend on the static charges. Hence, our transformation is valid for a given sector on the Hilbert space, or in other words, constructed to fit a the sector of interest. Thanks to the superselection of static charges, there is no point in discussing more than a single sector, and this is the point where we make an explicit choice of the sector, discarding all other sectors henceforth.

By construction, the matter qubits are completely decoupled in the transformed state, as the transformed Gauss' law (apply $\mathcal{U}^{\paren{2}}$ to \fulleqref{eqn:Gauss_with_projectors}) is:
\begin{equation} \label{eqn:gauss_law_method2}
	\sigma^z_n\left|\psi^{(2)}\right\rangle=-\left|\psi^{(2)}\right\rangle, \hspace{10pt} \forall n,
\end{equation}
- transformed physical states are ones in which all matter qubits are in the $\sigma^z_n=-1$ state.
In other words, we started with a state with gauge fields and matter, satisfying the local Gauss' law constraints, and ended up with a state where the gauge fields and matter are decoupled, and the local constraints are satisfied by the matter degrees of freedom alone. Originally, the Hilbert space was divided into dynamically disconnected sectors given by Gauss' law, and now the sectors are of the decoupled matter alone ($\left[H^{(2)},\sigma^z_n\right]=0$ $\forall n$). For this reason, in the beginning, while being constrained, we could not simply discard the matter degrees of freedom, now it possible to do so thanks to the decoupling. 

We can thus restrict ourselves to the sector where all the matter spins point down. They are not affected by the dynamics, and hence they do not have to be simulated. 
Formally, if we define by $\ket{\text{out}} \in \mathcal{H}_\text{m}^{(2)}$ the matter-state for which $\sigma^z_n=-1$  $\forall n$,
our relevant quantum simulation Hamiltonian will be 
\begin{equation} \label{eqn:H2_tilde}
    \tilde{H}^{(2)}=\Qmatrixelement{\text{out}}{H^{(2)}}{\text{out}}.
\end{equation}
 $\tilde{H}^{(2)}$ acts only on field (link) qubits states
 \begin{equation}
     \ket{\tilde{\psi}^{(2)}}=\Qbraket{\text{out}}{\psi^{\paren{2}}}\in \mathcal{H}^{\paren{2}}_{\text{g}},     
 \end{equation}
and describes the same physics as the original $H$ in a specific chosen charge-sector defined by the choice of $\parcr{\varepsilon_n}$. 
We thus arrive at a theory in a much smaller Hilbert space, but with no local constraints, containing only the relevant part of the spectrum (as can be seen in Fig. \ref{Fig3}). This is the result of combining a unitary transformation (preserving the spectrum) and a projection (which keeps only the relevant part of it). By simply plugging different static charges to the definitions of the projectors and the transformation, one can obtain a similar result for any other sector.

Importantly, the choice of matter sector (on our case - all matter spins pointing down) is not completely orthogonal to the choice of charge-sector $\parcr{\varepsilon_n}$, and one has to check for consistency with the \emph{global} charge symmetry, 

\begin{equation} \label{eqn:global_charge_symmetry}
    e^{i\pi\sum_n N_n}\ket{\psi} = e^{i\pi q}\ket{\psi},
\end{equation}
where $q=\sum_n q_n$. Since the sign of the right hand side is determined by the static charge sector and the left hand side by the fermionic parity sector, the two choices have to be made such that \fulleqref{eqn:global_charge_symmetry} is fulfilled. In practice this means that for charge sectors with an odd $q$, one would have to use a slightly different matter sector instead of the one we use here (for example - the first matter spin in the chain points up, and all the others point down), and the decoupling operation $\mathcal{U}_n$ would have to be changed accordingly.

Applying this procedure to $H^{\paren{1}}$ (\fulleqref{eqn:H_1_1d}),  one finds that the electric term remains unchanged, the mass term becomes the local two-body interaction: 
\begin{equation}
    \tilde{H}^{(2)}_{\text{m}} = -\frac{m}{2}\underset{n}{\sum}\paren{-1}^n \varepsilon_n Z_{n}Z_{n+1},    
\end{equation}
and the interaction term takes the form 
\begin{equation} \label{eqn:H_GM_tilde_2}
    \tilde{H}^{(2)}_{\text{GM}}= 
-\frac{J}{2}\underset{n}{\sum}\left(-\varepsilon_n\right)Y_n\left(1+Z_{n-1}Z_{n+1}\right).
\end{equation}
This procedure (applying $\mathcal{U}^{(2)}$ to $H^\text{(1)}$ and projecting on $\ket{\text{out}}$) is described in more detail in Appendix A.

At this point we focus an the specific charge-sector with $\varepsilon_n=\paren{-1}^n$ (chosen to include the "Dirac-sea" state). This is consistent with our matter-sector choice only when $L$ is an integer multiple of $4$, so from here on we restrict ourselves to this case. Note that the procedure can be easily altered to fit the other even $L$ case instead (by choosing a different matter-sector, and changing $\mathcal{U}_n$ accordingly as explained above).
As a result, the mass term simplifies, but $ \tilde{H}^{(2)}_{\text{GM}}$ still has an alternating sign $\paren{-\varepsilon_n}=\paren{-1}^{n+1}$. This is not a problem, but for the sake of elegance we make one extra step, using 
\begin{equation}
\mathcal{V} =\mathcal{V}^{\dagger}= \overset{L/2}{\underset{n=1}{\prod}}Z_{2n},
\end{equation}
to finally obtain:
\begin{equation} \label{eqn:method_2_final}
	\hat{H}=\mathcal{V}\tilde{H}^{(2)}\mathcal{V}=\hat{H}_{\text{E}} + \hat{H}_{\text{m}} + \hat{H}_{\text{GM}},
\end{equation}
where
\begin{align}
    \hat{H}_{\text{E}} &= -\underset{n}{\sum}\left( h Z_n + \frac{J}{2}Y_n\right),\\  \label{eqn:method_2_final_He}
\hat{H}_{\text{m}} &= \tilde{H}^{(2)}_{\text{m}} = -\frac{m}{2}\underset{n}{\sum} Z_{n}Z_{n+1},    \\
		\hat{H}_{\text{GM}}&= 
-\frac{J}{2}\underset{n}{\sum}Z_{n-1}Y_nZ_{n+1}, \label{eqn:method_2_final_Hgm}   
\end{align}
and we have re-defined the interaction and electric parts such that the former includes only three-qubit interactions and the latter has all the single qubit terms (including those that came from the original interaction part).

To change to open boundary conditions one can use almost the same expressions, but remember to sum over the sites $0\le n \le L-1$ for $\hat{H}_{\text{m}}$, and over the links $0\le n\le L-2$ for $\hat{H}_\text{E}$ and $\hat{H}_\text{GM}$. Then one has to make the substitution $Z_{-1}=Z_{L-1}=1$ which can be thought of as placing two additional links at the boundaries, with fixed field values. 
The result is the addition of boundary terms that are simpler than the bulk terms (single-qubit instead of two-qubit terms, and two-qubit instead of three-qubit terms).

In both cases, we now have an $L-1$ link-qubits Hamiltonian (Fig. \ref{Fig2}(c)), acting on states $\left|\hat{\psi}\right\rangle = \mathcal{V}\ket{\tilde{\psi}^{\paren{2}}}$ without constraints, global or local: again, an exponential reduction of the Hilbert space (see Fig. \ref{Fig3}(a,c) for a comparison of the spectra of $H$ and $\hat{H}$). However, in contrast to the standard method of $H^{\text{(0)}}$ (section \ref{method_0}), we now have a local Hamiltonian, and the procedure is generalizable to higher dimensions (see section \ref{sec:2d_theory}).

\subsection{From Hamiltonian to quantum simulation} \label{trotterization_etc}
Time evolution with respect to $\hat{H}$ is readily implemented using Trotterization. $\Omega_{\text{E}}\paren{\epsilon} \equiv e^{-i\epsilon \hat{H}_{\text{E}}}$ (where $\epsilon$ is the length of a Trotter step) can  be implemented in an analogue fashion (that is, as a whole) during the Trotter step; on the other hand, in a more digital approach, it can be decomposed into a product of local, commuting single qubit rotations,
\begin{equation}\label{eqn:trotter_He_1d}
\Omega_{\text{E}}\paren{\epsilon}  = \underset{n}{\prod} \exp{\parsq{-i r \epsilon \left(\cos\theta Z_n + \sin\theta Y_n\right)}},
\end{equation}
where  $r=\sqrt{h^2 + J^2/4}$, and $\cos\theta = -h/r$ and $\sin\theta = -J/2r$ define the axis of rotation in the $ZY$ plane. It is very likely that any simulating platform will be able to run all these gates in parallel, and even if not, it should be possible to do it in a finite number of steps where several qubits are rotated in parallel. Thus, for all practical purposes one can assume that $\Omega_{\text{E}}$ is implemented in an analogue way.
 
A similar argument holds for $\Omega_{\text{m}}\paren{\epsilon} \equiv e^{-i\epsilon \hat{H}_{\text{m}}}$. Here, instead of local terms we have two-body $ZZ$ interactions of nearest neighbours (and local rotations for the ends of the open system) which mutually commute, and may be implemented either together (analogically -- note that $\hat{H}_{\text{m}}$ is nothing but a simple Ising Hamiltonian) or sequentially with a small number of steps (since some of the constituent operations can be run in parallel, depending on the simulating platform).

Finally, $\Omega_{\text{GM}}\paren{\epsilon} \equiv e^{-i\epsilon \hat{H}_{\text{GM}}}$ would be more challenging for most simulation platforms, since it involves three-body interactions which are not natural for them. To implement it, we use the conventional controlled-Z gate:
\begin{equation} \label{eqn:CZ_def}
\begin{split}
    U^{\text{CZ}}_n &= \frac{1}{2}\left(1+Z_n+Z_{n+1}-Z_nZ_{n+1}\right)\\
    &=\exp\left(\frac{i\pi}{4}\left(1-Z_n\right)\left(1-Z_{n+1}\right)\right),
\end{split}
\end{equation}
 which obeys 
 \begin{equation} \label{eqn:CZ_property}
 \begin{split}
     &U^\text{CZ}_n Y_n U^\text{CZ}_n = Y_n Z_{n+1}\\
     &U^\text{CZ}_{n-1} Y_n U^\text{CZ}_{n-1} = Z_{n-1} Y_n.
 \end{split}
 \end{equation}
 Defining $U^{\text{CZ}}=\underset{n}{\prod}U^{\text{CZ}}_n $, it follows from \fulleqref{eqn:CZ_property} that
\begin{equation} \label{eqn:CZ_trotterization}
    \Omega_{\text{GM}}\paren{\epsilon}= U^{\text{CZ}}U_Y\paren{\epsilon} U^{\text{CZ}},
\end{equation}
where $U_Y\paren{\epsilon} = \exp\left(i\epsilon J\underset{n}{\sum}Y_n/2\right)\equiv \exp\left(-i\epsilon H_Y\right)$
  which, again, can be run either in parallel or sequentially, depending on technological constraints of the simulating platform. In either case, it can be done with a finite number of steps, independent of $L$, and we conclude that the entire algorithm runs in $O(1)$ time.
  
  The only remaining task is to choose the order of the three unitaries out of which a Trotter step is built. A Trotter error analysis, which is given in Appendix B, shows that the optimal ordering is
  \begin{equation} \label{eqn:trotter_order}
  	e^{-i\hat{H}t}\approx\left[\Omega_{\text{GM}}\paren{\epsilon}\Omega_{\text{m}}\paren{\epsilon}\Omega_{\text{E}}\paren{\epsilon}\right]^{\mathcal{N}}
  \end{equation}
which can be applied as a recipe for a fully digital quantum simulation of the model.

The exponential form of the CZ operation can be used to construct a hybrid analogue-digital simulation: First, use \fulleqref{eqn:CZ_def} to express $U^\text{CZ}$ as $ \exp{\paren{-i\epsilon H_Z}}$, where (up to an irrelevant constant)
 \begin{equation}
    H_Z =  -\frac{\pi}{4\epsilon}\underset{n}{\sum}Z_nZ_{n+1}+
\frac{\pi}{2\epsilon}\underset{n=2}{\sum}Z_n.     
 \end{equation}
Since $H_Z$ and $\hat{H}_{\text{m}}$ not only commute, but also have a very similar functional form,
we can define
\begin{equation}
    \hat{H}_Z = -\left(\frac{m}{2}+\frac{\pi}{4\epsilon}\right)\underset{n}{\sum}Z_n Z_{n+1}+
\frac{\pi}{2\epsilon}\underset{n=2}{\sum}Z_n,    
\end{equation}
which is a simple Ising Hamiltonian with a longitudinal field. Then we can obtain our single Trotter step using a sequence in which we switch on and off four analogue Hamiltonians:
\begin{equation} \label{eqn:hybrid}
    e^{-i\hat{H}t}\approx\left(	e^{-i\epsilon H_Z}e^{-i\epsilon H_Y}e^{-i\epsilon \hat{H}_Z}e^{-i\epsilon \hat{H}_{\text{E}}}\right)^{\mathcal{N}}.
\end{equation}

After simulating time evolution (either in the hybrid or in the fully digital way), we have to be able to measure observables from the original model. The relevant local observables are the electric field
\begin{equation} \label{eqn:E_def}
    E_n=\frac{1}{2}\paren{1-Z_n}
\end{equation}
 on the links, and the number operator $N_n=\psi_n^{\dagger}\psi_n$ on the sites. To measure these, we first have to check how they transform under our procedure: first with $\mathcal{U}^{\text{(1)}}$, then with $\mathcal{U}^{\text{(2)}}$, and finally projecting the matter state onto $\ket{\text{out}}$ and rotating with $\mathcal{V}$ (though in these cases $\mathcal{V}$ has no effect). It is easily verified that under this procedure $E_n$ and $N_n$ transform to: 
\begin{align}
    \label{eqn:field_E_transformed} \hat{E}_n &=E_n=\frac{1}{2}\paren{1-Z_n}\\  
    \hat{N}_n &=\frac{1}{2}\paren{1-\varepsilon_n S_n}.  \label{eqn:N_fermions_transformed}
\end{align} 
The field $E_n$ is unchanged and can therefore be obtained trivially from measuring the qubits in the computational basis, while for $N_n$ we have to measure the product $S_n =Z_{n-1} Z_n$, which is the parity of neighbouring qubits. 
The non-local gauge invariant observables are the mesonic strings, defined in \fulleqref{eqn:mesonic_string_general}. Measuring those is also possible within this scheme, but it is somewhat more involved and we show how to do it in Appendix C.

This concludes our construction for the one-dimensional case. Such a simulator would be useful for a broad range of tasks, e.g. adiabatic ground state preparation, or studying quenches, some of which are exemplified in the following section.

%%%%%%%%%%%%%%%%%%%%%%%%%%%%%%%% 1D EXPERIMENT %%%%%%%%%%%%%%%%%%%%%%%%%%%%%%%%%%%
\begin{figure*} 
    \includegraphics[width = \textwidth]{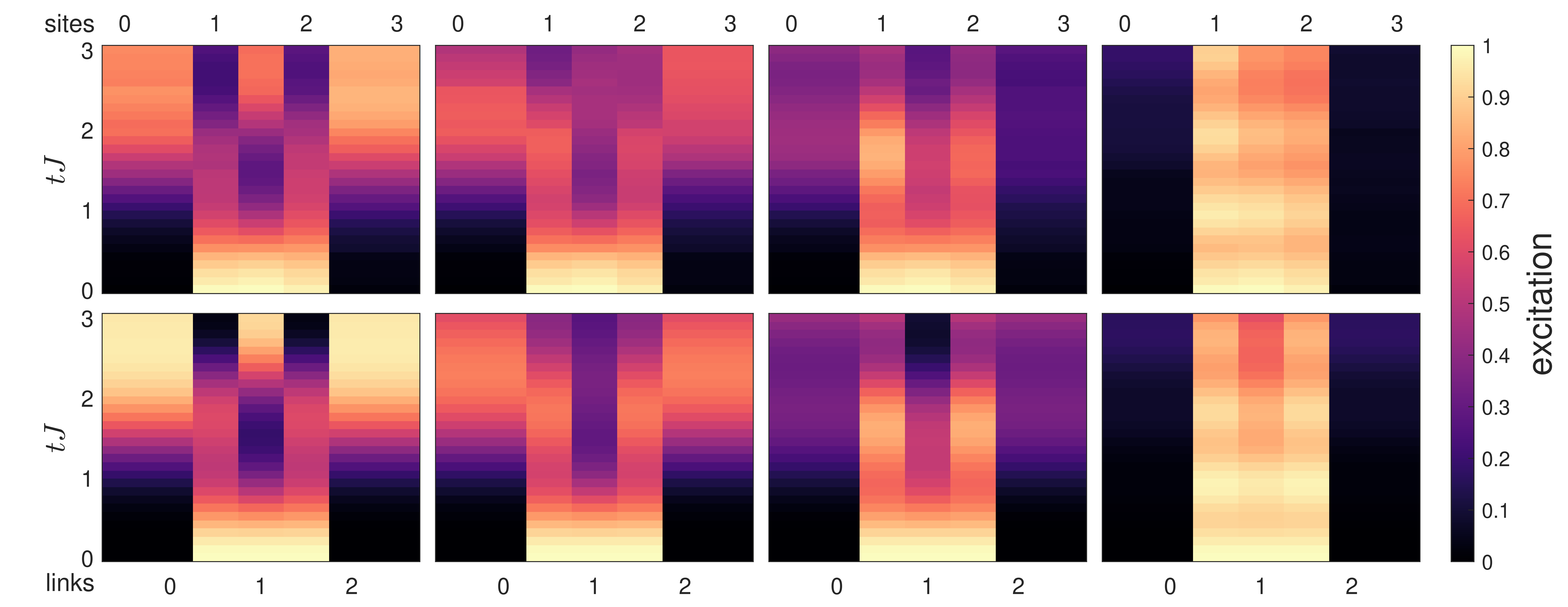}
    \centering
    \caption{Time evolution of the $1+1d$ model with $L=4$ sites, from an excitation of the middle ($n=1$) link. (top) Measurement on \emph{ibm-lagos} and (bottom) exact numerical solution, with different values of $h/J$ (left-to-right: 0.1, 0.5, 1, and 3). Plotted is the excitation with respect to the "Dirac-sea" state: that is, for the links we plot the field $\left\langle E_{n}\right\rangle$  for the even sites (0 and 2) we plot the number of fermions $\left\langle N_{n}\right\rangle$, and for the odd sites (1 and 3) we plot $\left\langle 1-N_{n}\right\rangle$  that can be thought of as the number of anti-particles. Confinement dynamics is observed for large $h/J$.  }
    \label{fig:1d_time_evolution}
\end{figure*}

\begin{figure} 
    \includegraphics[trim=22 25 22 0,clip,width = \columnwidth]{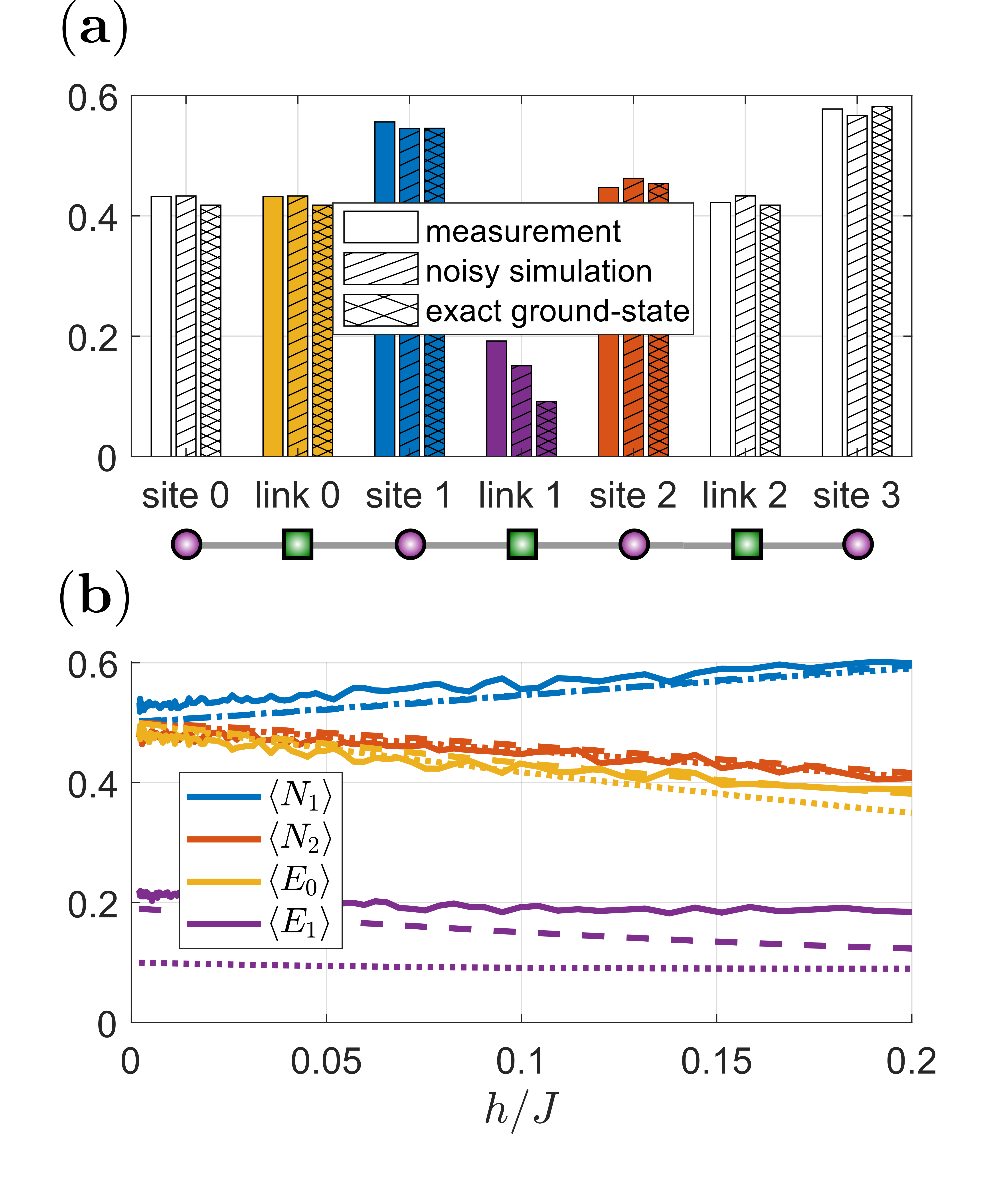}
    \centering
    \caption{Adiabatic ground state preparation experiment with $L=4$ sites. (a) Expectation values for the local observables ($N_{n}$  at the sites and $E_{n}$ on the links) for $h/J=0.1$. (b) A subset of the observables plotted against different values of $h/J$: measurement on \emph{ibmq-quito} (solid), noisy numerical simulation (dashed) and exact diagonalization (dotted).}
    \label{fig:1d_gs_vs_h}
\end{figure}

\section{Experimental implementation} \label{1d_experiment}
We implemented a proof-of-concept version of this quantum simulation proposal via the IBMQ platform. For that we focus on the $1+1d$ case with $m=0$ and open boundary conditions, in the $\varepsilon_n=\paren{-1}^n$ sector. This means that we have to implement the  $L-1$ qubits Hamiltonian:
\begin{equation} \label{eqn:sim_H_1D}
    \hat{H}=\hat{H}_{\text{E}} + \hat{H}_{\text{GM}}   
\end{equation}
with 
\begin{align}
    \hat{H}_{\text{E}} &= -\sum_{n=0}^{L-2}\left(h Z_n + \frac{J}{2}Y_n\right) \\
    -\frac{2}{J} \hat{H}_{\text{GM}} &= \sum_{n=1}^{L-3} Z_{n-1}Y_{n}Z_{n+1} + Y_0 Z_1 + Z_{L-3}Y_{L-2},
\end{align}
where the last two terms are boundary terms. The hybrid analogue-digital approach (\fulleqref{eqn:hybrid}) might possibly be implemented on those IBMQ devices that allow for direct pulse control, but this is beyond the scope of this work. Instead we follow the Trotterization procedure for a fully digital simulation, which can be summarized by Eq. \eqref{eqn:trotter_He_1d} and \eqref{eqn:CZ_trotterization} (importantly, these hold for the open boundary conditions Hamiltonian as well), and split the electric part in half to reduce the Trotter error, implementing:
\begin{equation}\label{eqn:exp_trotter_step}
    e^{-i\hat{H} t }\approx \parsq{\Omega_{\text{E}}\paren{\epsilon/2}  \Omega_{\text{GM}}\paren{\epsilon}
    \Omega_{\text{E}}\paren{\epsilon/2}}^{\mathcal{N}}.
\end{equation}

The operation $U^{\text{CZ}}$ (controlled-Z on all pairs of neighbouring qubits in the chain) that appears twice in $\Omega_{\text{GM}}\paren{\epsilon}$ has to be implemented in two steps (one for the even pairs and another for the odd pairs). This means that each Trotter step can be implemented with $4$ two-qubit gate steps, and $2$ single-qubit rotation steps. We emphasize again that these numbers do not depend on $L$. Converting from CZ gates and general single-qubit rotations to the native gates of the IBMQ devices (CNOT, X, $\sqrt{\text{X}}$ and virtual Z gates) costs in additional $4$ single-qubit steps. 

Typical IBMQ qubits have coherence times on the order of 100 microseconds, and native single-qubit gates can be implemented within 35ns. Two-qubit (CNOT) gates however, are implemented with via the cross-resonance approach \cite{Paraoanu_cross_resonance_2006,Rigetti_Devoret_cross_resonance_2010} and typically take between 300-500ns each. Assuming we want the computation to complete within $\sim10\%$ of the coherence time, this restricts us to about $5$ Trotter steps in total (about 20 native two-qubit steps and 30 native single qubit steps where each native step acts on the entire chain). This poses a limitation on the possible computations. For example: when implementing adiabatic ground-state preparation, the adiabaticity condition cannot be fulfilled for some regions in parameter space, resulting in poor fidelities. Nevertheless, it is important to remember that faster or more coherent hardware does exist, and state-of-the-art technology already allows for an order of magnitude improvement in the coherent Trotter depth. The rather strict requirement of completing the experiment within $10\%$ of the coherence time is an empirically (and numerically) verified heuristic that seemed to optimize the Trotter error against decoherence errors in most of our experiments. However, it has been shown that error mitigation techniques like ZNE (which we did not implement here) allow for meaningful evaluation of observables even when a larger degree of decoherence noise is allowed in the experiment \cite{Giurgica-Tiron_ZNE_2020}.

One of the most significant advantages of quantum simulation is the possibility to simulate time evolution. Importantly, the limitation of 5 trotter steps does not translate to a limitation on the temporal resolution, since one can directly control the size of each step (which translates to an angle of rotation in a single-qubit gate). Practically this means that we have to choose the number (between 1 and 6 in this case) and the length (between $0.4/J$ and $0.5/J$) of the Trotter steps to fit each desired simulated evolution time.
For our demonstration we initialize the $L=4$ chain with an excitation in one of the qubits: this corresponds to an excitation of the field on the relevant link, as well as a change in the sites connected to it to accommodate the original gauge constraints. Then we evolve it in time and measure the local observables ($E_n$ on the links and $N_n$ on the sites) as a function of the evolution time. The measurement is averaged over 20000 to 30000 shots such that the readout error is insignificant. We observe (\figref{fig:1d_time_evolution}) that for small values of $h/J$ the initial excitation diffuses to the neighboring sites and links, while for large $h/J$ it remains confined. With only $4$ sites, we cannot claim to having observed a phase transition, however this is still a non-trivial physical feature of the model that our quantum simulation captures using only $3$ qubits and a few tens of noisy gates.

Quantum simulation can also be used to investigate non-trivial ground-states via adiabatic ground state preparation. For example, since the ground state of the $J=0$ Hamiltonian is trivial (all qubits are at $\left|0\right\rangle$, which corresponds to the "Dirac-sea" state of the original model), by running a time evolution experiment while increasing $J$ from zero with each step, we can measure the ground state for a finite $J$.

Motivated by recent work on scaling phenomena near the $h=0$ transition \cite{Frank_Chandrasekharan_emegrence_Z2_2020}, we chose to implement the opposite (increasing $h$ adiabatically) for the purpose of our proof-of-concept demonstration. Initializing the $h=0$ ground state is not as trivial as the $J=0$ ground state, but there is a simple shallow circuit that initializes the ground state of $\hat{H}_{\text{GM}}$ (that is, only the term that is interacting for the qubits, rather then the interaction term of the original model). This circuit is straightforward to derive based on \fulleqref{eqn:CZ_property}.

After this initialization we proceed by adiabatically increasing the non-interactive terms simultaneously, and arrive at the desired finite $h$ ground state. This scheme was implemented for $L=4$ sites and the results are summarized in \figref{fig:1d_gs_vs_h}, showing good agreement with the exact solution and with a noisy numerical simulation, implemented on Python via the Qiskit-Aer package. For transparency, we used a custom noise model that includes only energy-relaxation and dephasing channels, with $T_1$, $T_2$ for each qubit and duration for each gate as reported by IBMQ.

Thus, we can probe the ground states of both the small $h$ and the small $J$ regimes. Intermediate regimes are more challenging on the IBMQ devices due to the aforementioned limitation on the total number of Trotter steps, but we show numerically (\figref{fig:numerics}) that reasonable fidelities can be expected with current technology. This is discussed further in section \ref{discussion_and_summary}.

%%%%%%%%%%%%%%%%%%%%%%%%%%%%%%%% 2D THEORY %%%%%%%%%%%%%%%%%%%%%%%%%%%%%%%%%%%
\section{Generalization to two spatial dimensions} \label{sec:2d_theory}
As we showed in section \ref{method_0}, the traditional methods that treat the Hilbert space redundancy and the problem of simulating fermions cannot be extended beyond $d=1$. The reason for that is that the Jordan Wigner transformation assumes an order over the sites, which in $d>1$  would have to be defined in an arbitrary way that is highly non-local and does not respect the lattice geometry. This is possible to in principle but extremely impractical. Even worse - the construction of $\mathcal{U}^{(0)}$ relies an the existence of a well-defined solution (\fulleqref{eqn:gauss_solution_m0}) of the constraints for the gauge-field, which is not available in $d>1$.
In contrast, our procedure is completely local, and relies on a unique solution of the constraints for the matter, which is available in any dimension. We demonstrate it here for $\mathbb{Z}_2$ with  $d=2$, in the charge sector defined by a choice of signs 
\begin{equation}
\varepsilon\paren{\mathbf{x}}=e^{i\pi q\left(\mathbf{x}\right)}.
\end{equation}

First, consider the hard-core bosonic formulation of the model. Applying the procedure of \fullcite{zohar_eliminating_2018} to the Hamiltonian \eqref{eqn:LGT_general_ham} at $d=2$, we get (since the terms get rather complicated in terms of coordinates and directions, we show it graphically):

\begin{widetext}
\begin{equation}
	\begin{aligned}
&H^{(1)} = -h\underset{\mathbf{x},i}{\sum} Z\left(\mathbf{x},i\right)  +m\underset{\mathbf{x}}{\sum}\left(-1\right)^{x_1+x_2}\sigma^z\left(\mathbf{x}\right)
\\&-b\underset{p}{\sum}\left[
\vcenter{\hbox{\includegraphics[scale=0.54]{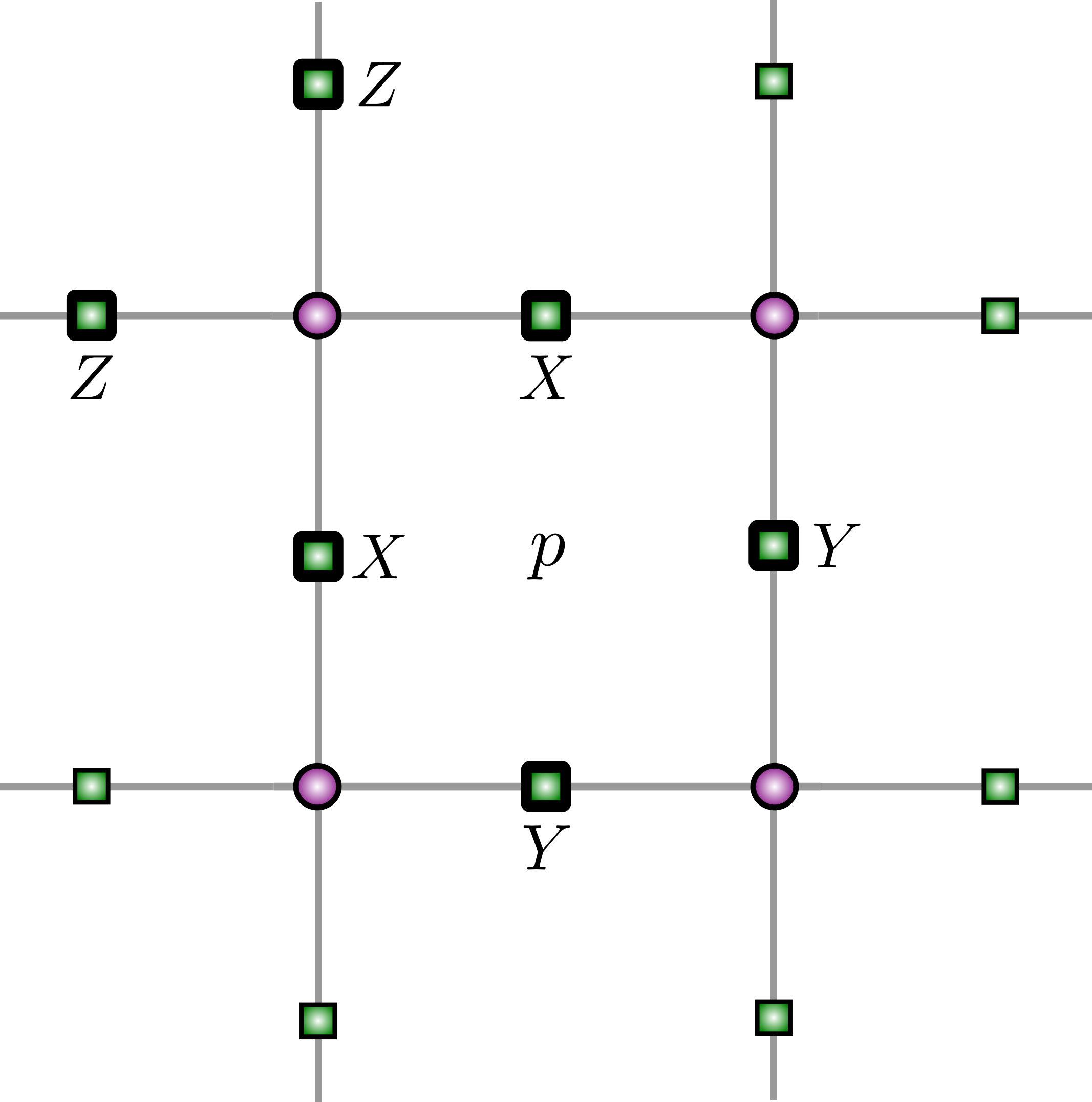}}}
\right]
+iJ\underset{\mathbf{x}}{\sum}
\left[
\vcenter{\hbox{\includegraphics[scale=0.54]{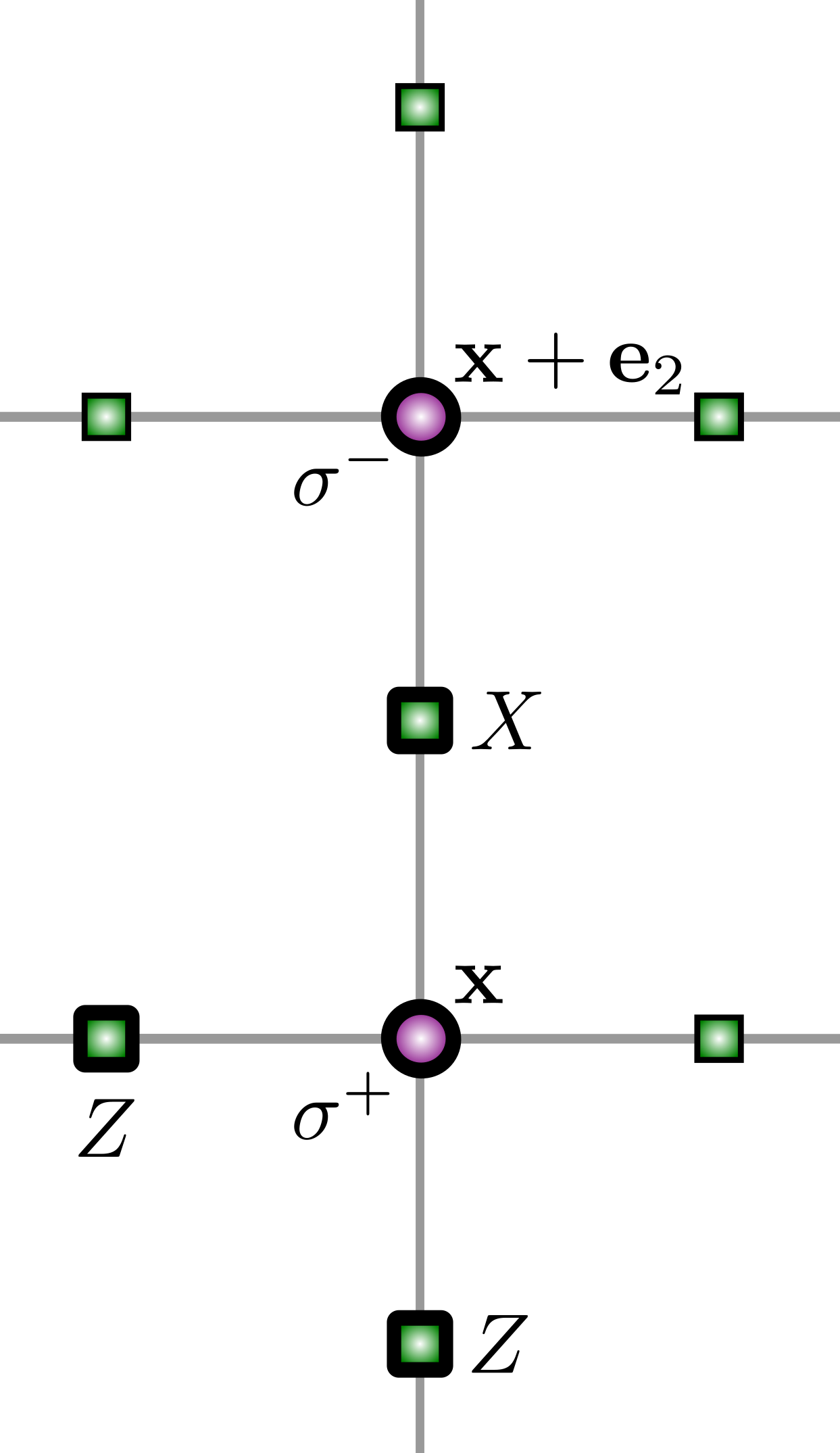}}} -\text{h.c.}
\right] 
+iJ\underset{\mathbf{x}}{\sum}\left[
\vcenter{\hbox{\includegraphics[scale=0.54]{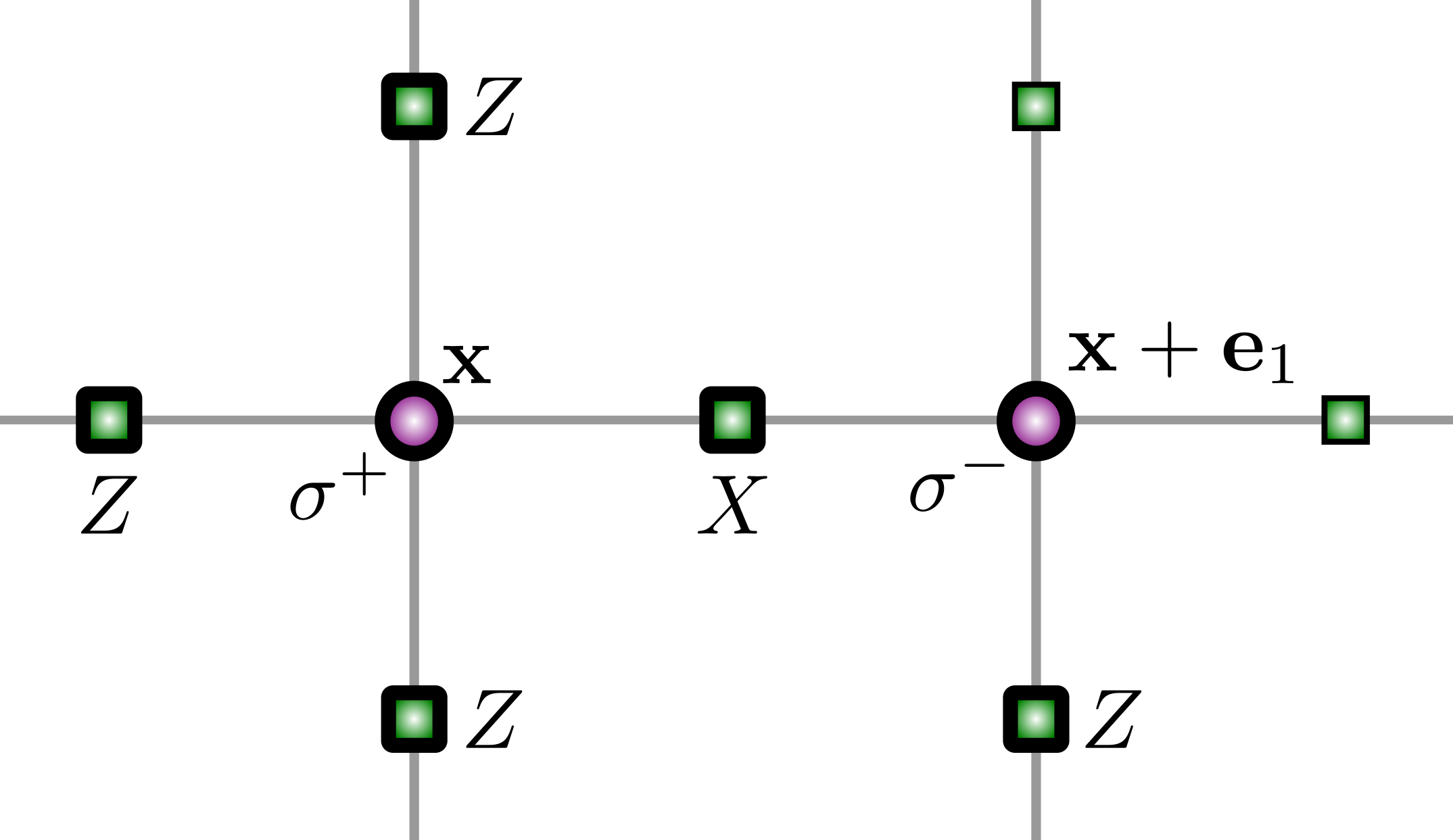}}} -\text{h.c.}
\right].
    \end{aligned}
\end{equation}
\end{widetext}

Gauss' law is very similar to that of \fulleqref{eqn:gauss_law_method1}: 
\begin{equation}
	\sigma^z\left(\mathbf{x}\right)\left|\psi^{(1)}\right\rangle = 
	-\varepsilon\paren{\mathbf{x}}S\left(\mathbf{x}\right)\left|\psi^{(1)}\right\rangle, \quad\forall \mathbf{x},
	\label{Gauss12d}
\end{equation}
with $S\left(\mathbf{x}\right)$ as defined in \fulleqref{gtrans}, completely analogous to $S_n$ in $d=1$. Here we see again that when treating Gauss' law as an equation for the matter ($\sigma^z(\mathbf{x})$) rather then for the field, it is explicitly solved, and extending to $d>1$ does not change that.

Therefore we can similarly define
\begin{equation}
P^{\pm}\left(\mathbf{x}\right) = \frac{1}{2}\left(1\mp\varepsilon\paren{\mathbf{x}} S\left(\mathbf{x}\right)\right),
\end{equation}
and rewrite Gauss' law as
\begin{equation}
	\left(P^{+}\left(\mathbf{x}\right)-P^{-}\left(\mathbf{x}\right)\right)\left|\psi^{(1)}\right\rangle = \sigma^z\left(\mathbf{x}\right)\left|\psi^{(1)}\right\rangle.
\end{equation}
The local controlled unitaries are defined the same way:
\begin{equation}
\mathcal{U}\left(\mathbf{x}\right) = P^{+}\left(\mathbf{x}\right) \sigma^x\left(\mathbf{x}\right) + P^-\left(\mathbf{x}\right),
\end{equation}
and since they all commute we can safely define $\mathcal{U}^{(2)} = \underset{\mathbf{x}}{\prod}\mathcal{U}\left(\mathbf{x}\right)$, from which the decoupling of matter follows, in the form of the new constraints 
\begin{equation}
	\sigma^z\left(\mathbf{x}\right)\left|\psi^{(2)}\right\rangle=-\left|\psi^{(2)}\right\rangle, \quad\forall \mathbf{x}.
	\label{Gauss22d}
\end{equation}

From this, we can obtain the $d=2$ Hamiltonian in a similar manner; it will involve \emph{local} few-body interactions (since $H^{(1)}$ is local, and $\mathcal{U}^{(2)}$ is local) which can be implemented using the same digital or digital-analogue tools. The locality guarantees that the Trotter steps can be concluded with an $O(1)$ run-time, as in the $d=1$ case. 
To get an idea of the result, we again focus on the simple charge sector $\varepsilon\paren{\mathbf{x}}=\paren{-1}^{x_1 + x_2}$, and restrict ourselves to $J\in\mathbb{R}$, which allows for some simplification in the resulting expressions:
\begin{widetext}
\begin{equation} \label{eqn:method2_2D}
	\begin{aligned}
		 &\tilde{H}^{(2)} =-h\underset{\mathbf{x},i}{\sum} Z\left(\mathbf{x},i\right)
        -\frac{m}{2}\sum_{\mathbf{x}}S\paren{\mathbf{x}}
  -b\underset{p}{\sum}\left[
		\vcenter{\hbox{\includegraphics[scale=0.53]{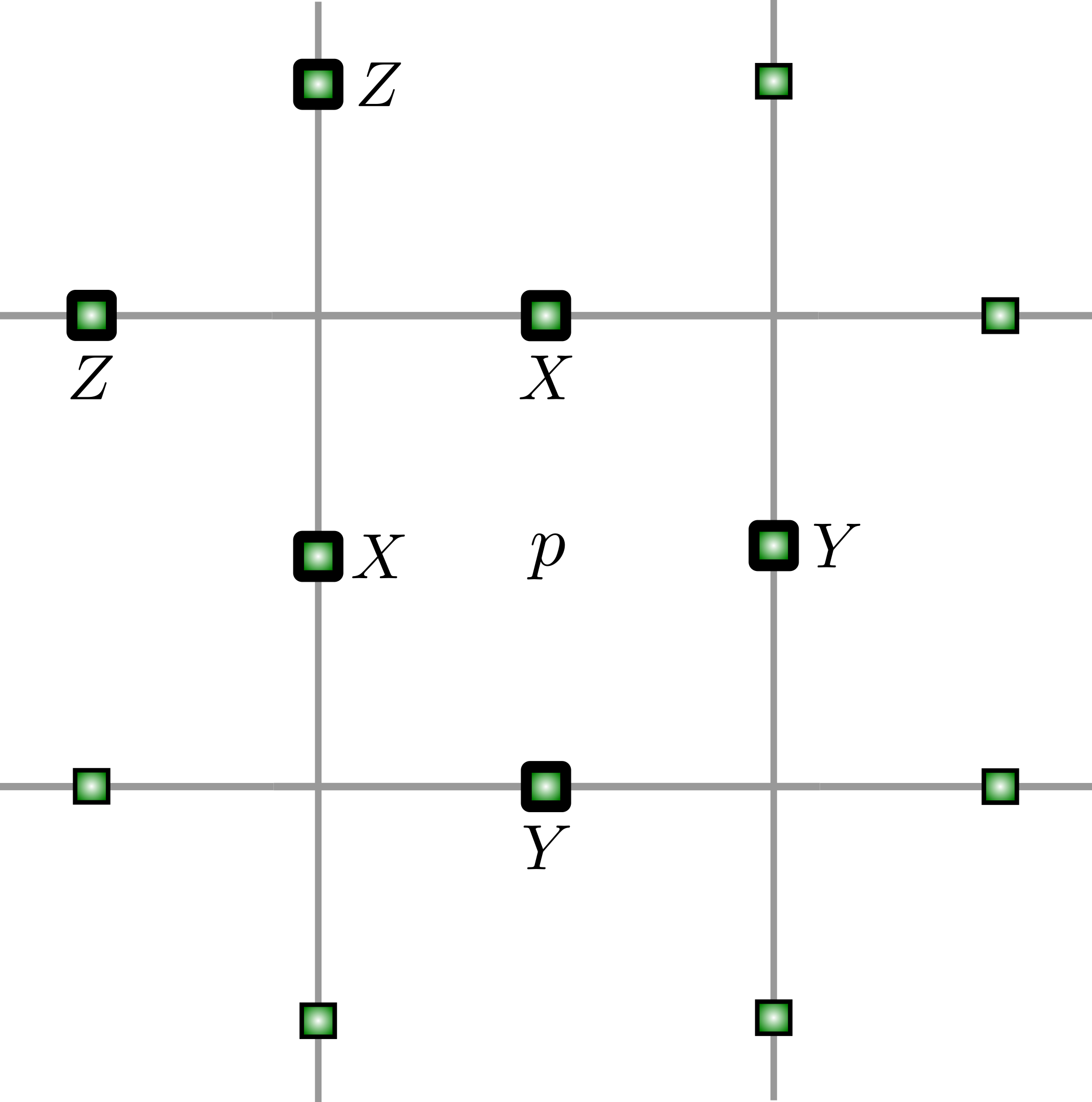}}}
		\right]
		+\left(-1\right)^{x_1+x_2}\frac{J}{2}\underset{\mathbf{x}}{\sum}\left[
		\vcenter{\hbox{\includegraphics[scale=0.53]{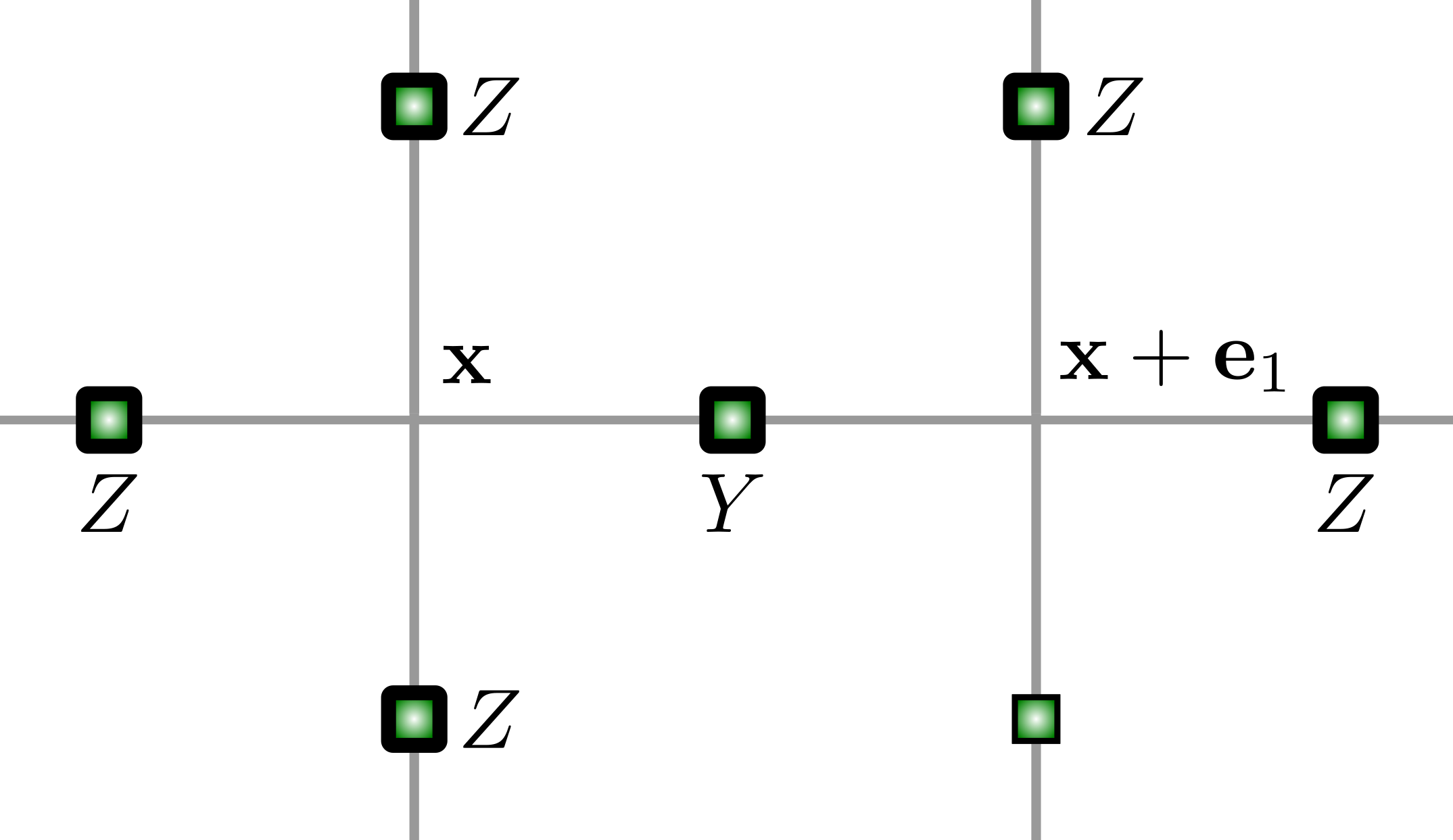}}}
		\right]
        \\&+\left(-1\right)^{x_1+x_2}\frac{J}{2}\underset{\mathbf{x}}{\sum}\left[
		\vcenter{\hbox{\includegraphics[scale=0.53]{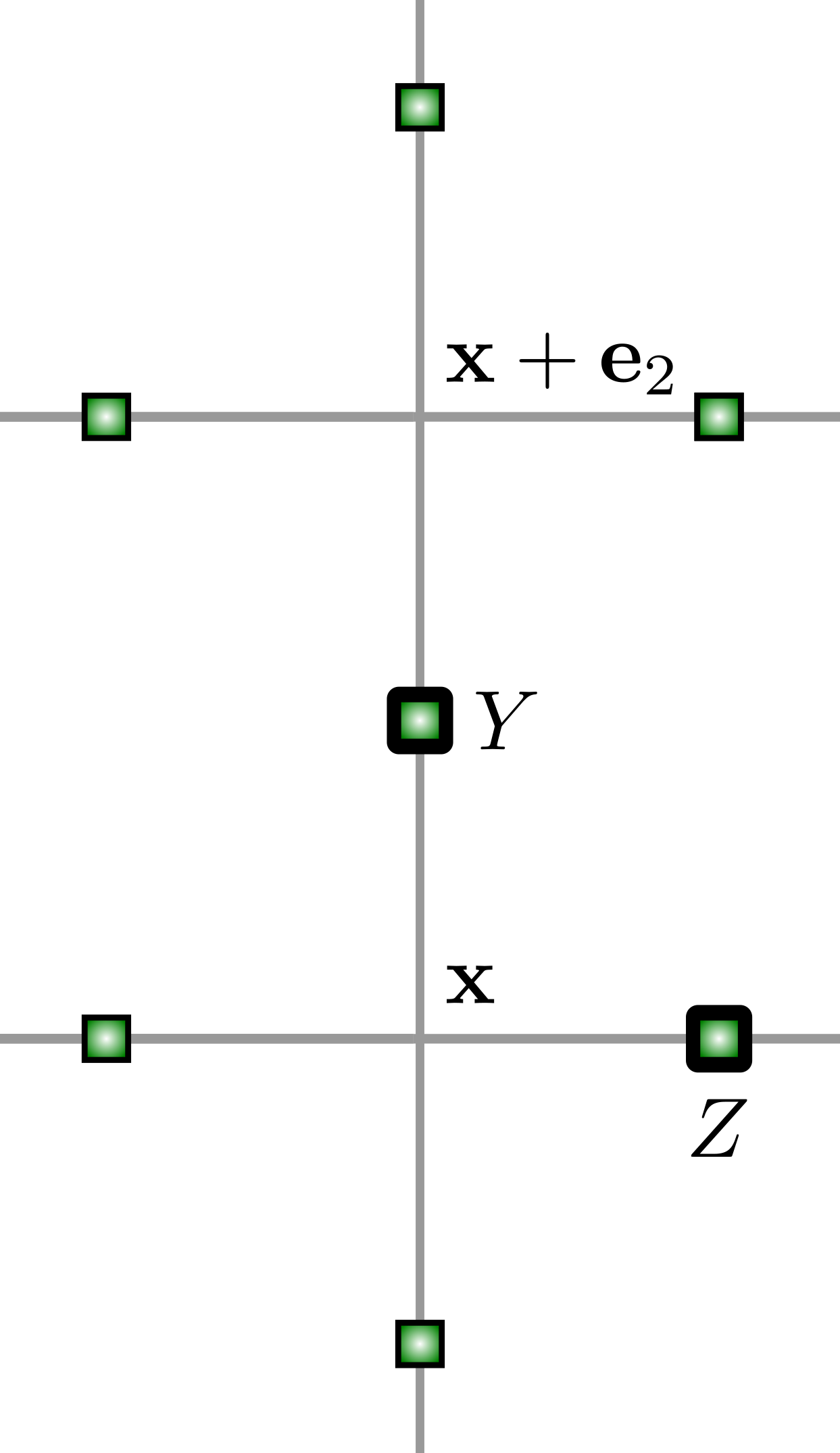}}}
		\right] 
		+\left(-1\right)^{x_1+x_2}\frac{J}{2}\underset{\mathbf{x}}{\sum}\left[
		\vcenter{\hbox{\includegraphics[scale=0.53]{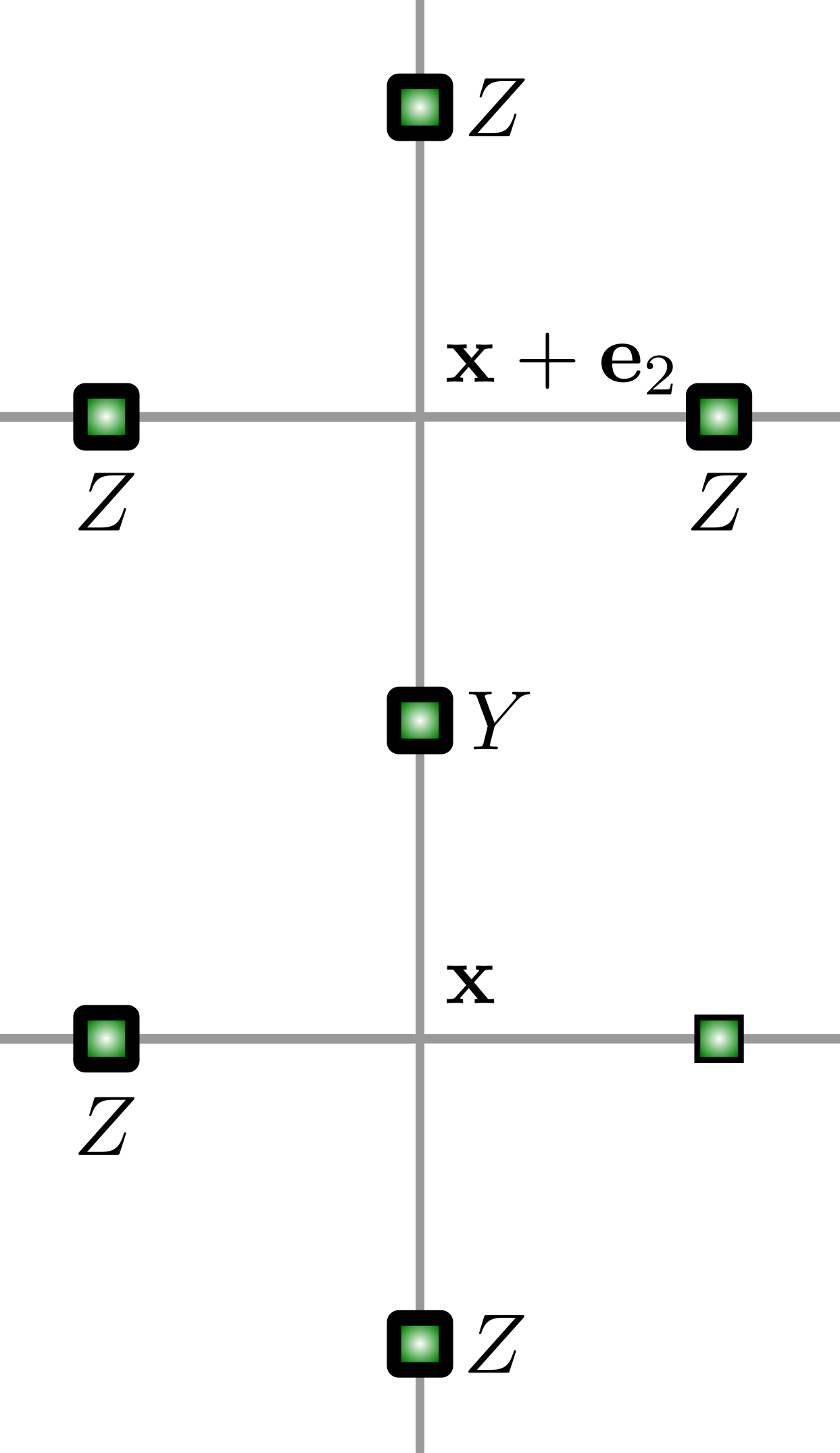}}}
		\right]
		+\left(-1\right)^{x_1+x_2}\frac{J}{2}\underset{\mathbf{x}}{\sum}\left[
		\vcenter{\hbox{\includegraphics[scale=0.53]{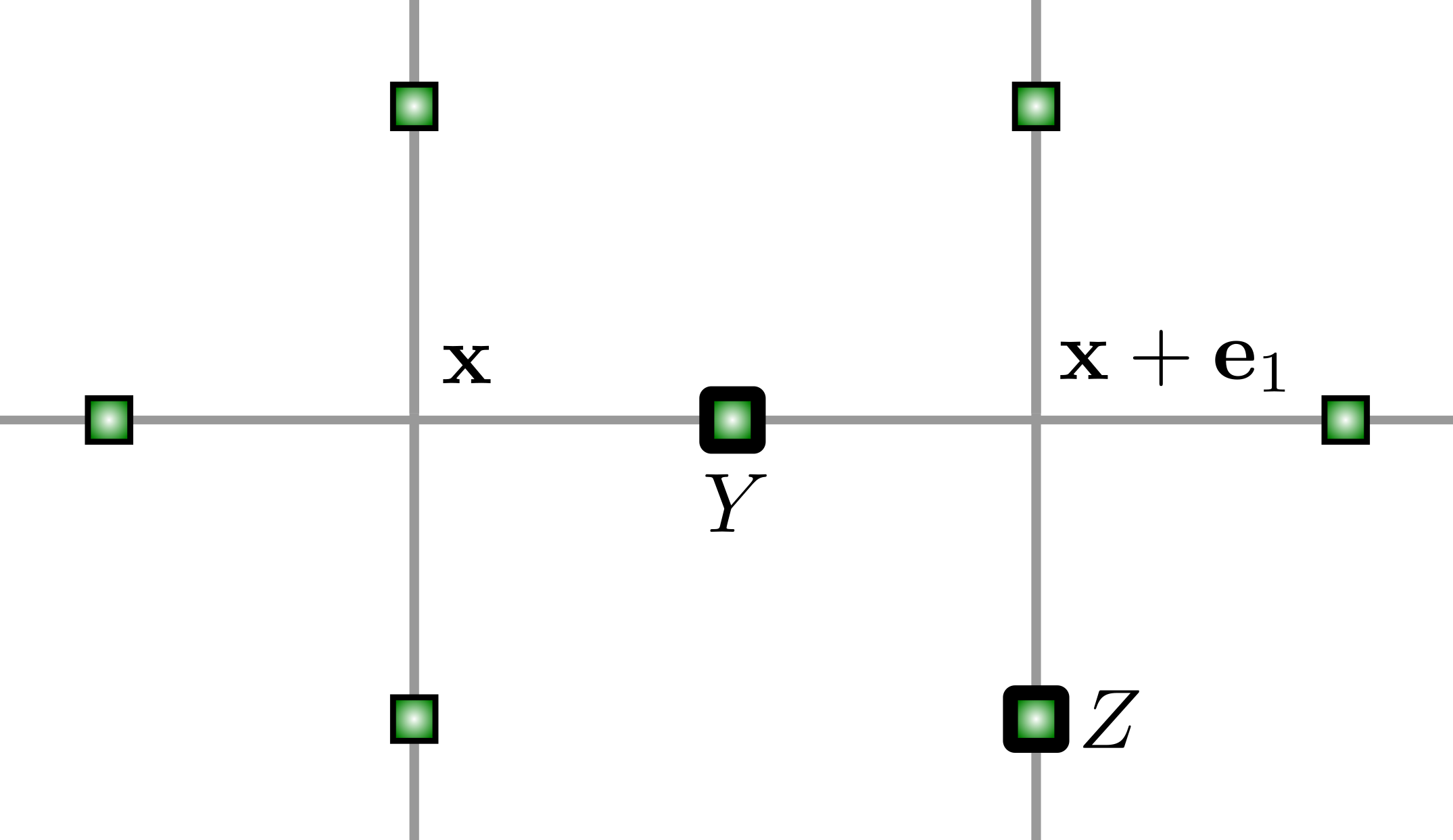}}}
		\right].
	\end{aligned}
\end{equation}
\end{widetext}
Here, too, we can remove the staggering with a unitary $\mathcal{V}$ which acts with $Z$ on all the links emanating from sites for which $x_1+x_2$ is even, and get the simulation Hamiltonian:
\begin{equation} \label{eqn:final_H_2D}
   	\hat{H}=\mathcal{V}\tilde{H}^{(2)}\mathcal{V}, 
\end{equation}
which has the exact same terms as in \fulleqref{eqn:method2_2D}, but without the alternating signs in front of the $J/2$ terms.
This is, as expected, a  Hamiltonian involving local qubit interactions, which can indeed be simulated using the usual quantum simulation techniques, such as those used for $d=1$ in section \ref{trotterization_etc}.

Moreover, one can repeat the entire procedure in the same way for $d>2$:  $H^{(1)}$ will have a slightly different form, but nevertheless local, and this will be the only significant change. All the arguments and techniques from section \ref{trotterization_etc} remain valid, and one is able to construct Trotter steps with $O(1)$ runtime.

%%%%%%%%%%%%%%%%%%%%%%%%%%%%%%%% 2D EXPERIMENT %%%%%%%%%%%%%%%%%%%%%%%%%%%%%%%%%%%
\begin{figure} 
    \includegraphics[width = 0.85\columnwidth]{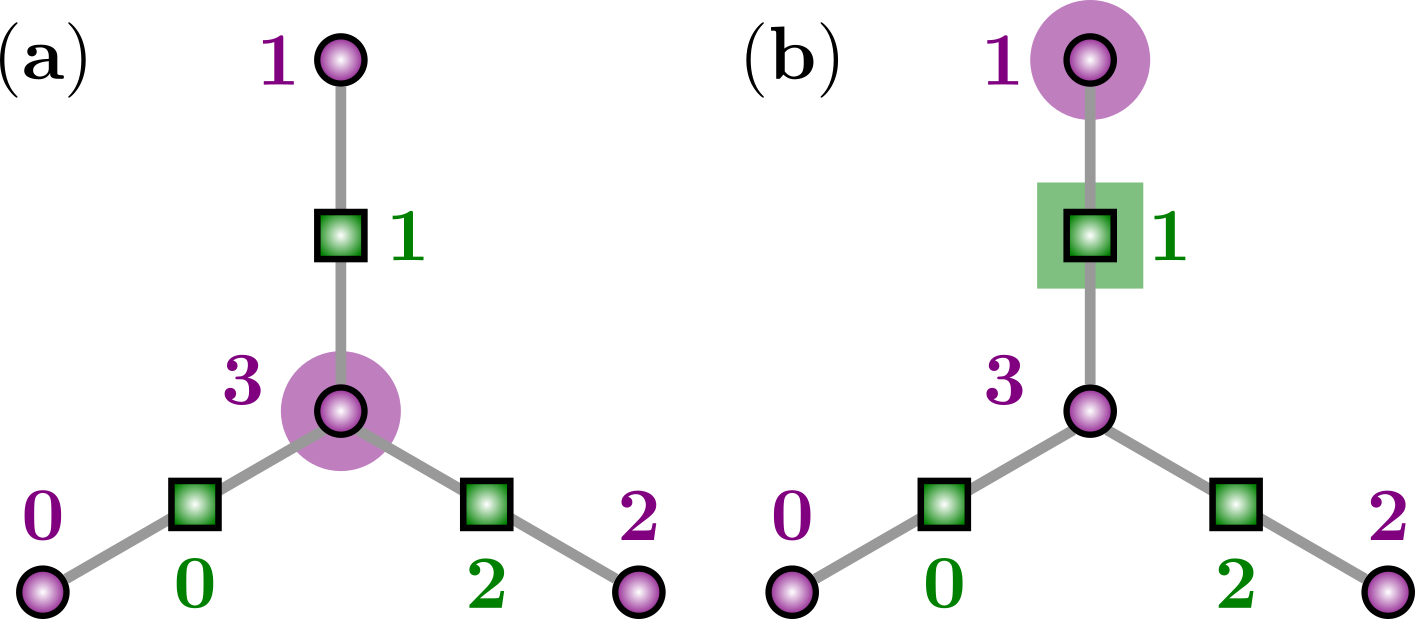}
    \centering
    \caption{The quasi two-dimensional system with four sites, and the indexing convention for the sites (in purple) and for the links (in green). The middle site ($n=3$) is considered an "odd" site in our chosen sector, which implies that (a) the $J=0$ ground-state is the one where $N_3=1$ and all other $N_n$ and $E_n$ equal zero (as indicated by the purple highlighting of the middle node). (b) The initial state of the time evolution experiment (\figref{fig:2d_time_evolution}), with excitation in qubit (link) 1, is the one where $N_1=E_1=1$ and all other $N_n$ and $E_n$ equal zero (as indicated by the highlighting of node 1 and link 1).
    }
    \label{fig:2d_layout}
\end{figure}
\begin{figure*} 
    \includegraphics[width = \textwidth]{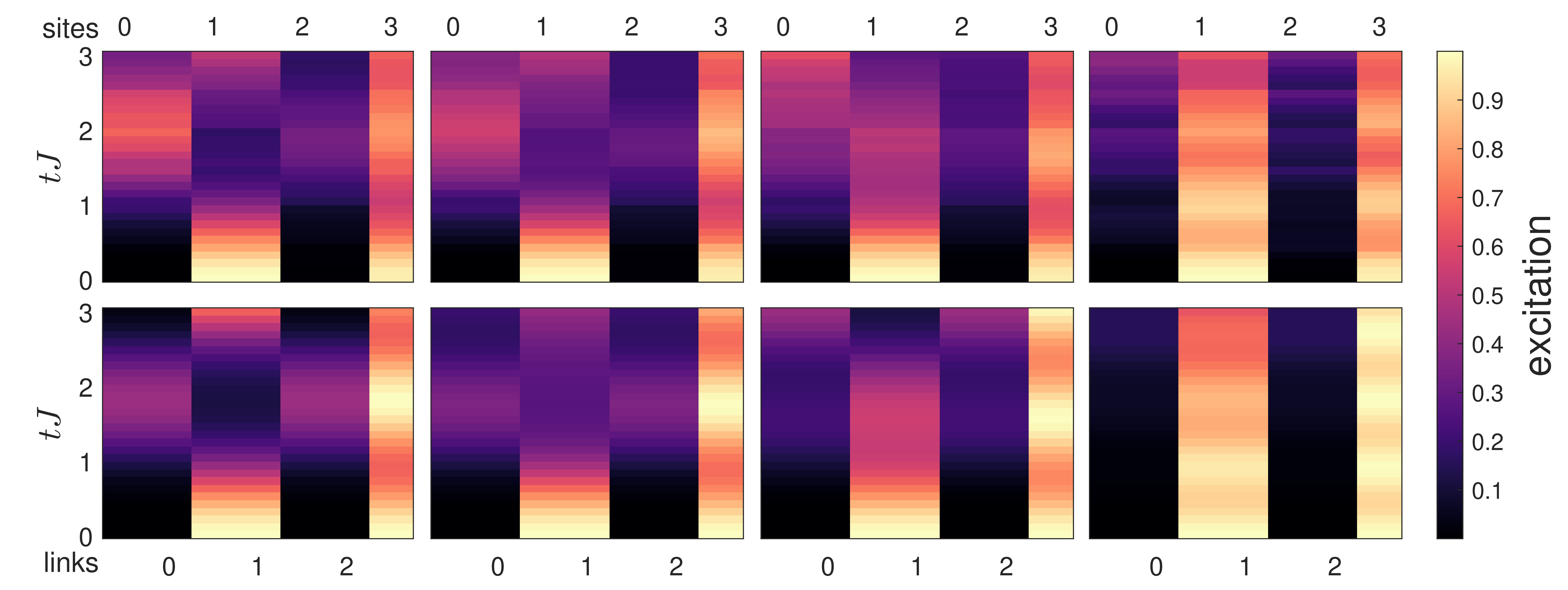}
    \centering
    \caption{ Time evolution experiment on the quasi two-dimensional model depicted in \figref{fig:2d_layout}. (top) Measurement on \emph{ibmq-lima} and (bottom) exact numerical solution, with different values of $h/J$ (left-to-right: 0.1, 0.5, 1, and 3). Plotted is the excitation with respect to the $J=0$ ground state (\figref{fig:2d_layout}(a) ): that is, for the links we plot the field $\left\langle E_{n}\right\rangle = \left\langle\frac{1}{2}\left(1-Z_{n}\right)\right\rangle$, for sites $n=0,1,2$ we plot $\left\langle N_{n}\right\rangle$  and for the middle site ($n=3$) we plot $\left\langle 1-N_{n}\right\rangle$  that can be thought of as the number of anti-particles. The initial state is the one where qubit 1 is excited, which corresponds to the original-model state shown in \figref{fig:2d_layout}(b). For large $h/J$ the initial state is more robust to the dynamics. 
    }
    \label{fig:2d_time_evolution}
\end{figure*}

\section{Experimental implementation of a quasi two-dimensional system} \label{2d_experiment}
In order to implement the $2+1d$ version of our proposal (\fulleqref{eqn:final_H_2D}) the qubits have to be organized on a square lattice. As this is not the case for any IBMQ machine, we implemented a quasi two dimensional version of the model with 4 sites as depicted in \figref{fig:2d_layout}, where the middle site is treated as an “odd” site for the purposes of staggering and choosing a superselection sector. This means that the Gauss' laws on the four sites are
\begin{equation}    
\begin{split} \label{eqn:barely_2d_gauss}
    &Z_n\ket{\psi} = e^{i\pi N_n}\ket{\psi}, \hspace{0.1\columnwidth} \text{for} \hspace{0.02\columnwidth} n=0,1,2,\\
    &Z_0Z_1Z_2\ket{\psi}= - e^{i\pi N_3}\ket{\psi},
\end{split}
\end{equation}
which implies that at $J=0$ the ground state is the one where $E_{n} =0$ and $ N_{n} =0$ for $n=0,1,2$, and $ N_{3} =1$). This toy-model is the simplest system where the standard approaches for eliminating the fermions fail due to the dimensionality and the connectivity. Assuming $m=0$ as in section \ref{1d_experiment} and following the matter elimination procedure, we find that the electric term $H_{\text{E}}$ does not change, and the interaction term $H_{\text{GM}}$ becomes:

\begin{equation} \label{eqn:sim_H_2D}
    -\frac{2}{J}\hat{H}_{\text{GM}}=\left(Y_{0}+Y_{2}\right)+\left(Z_{0}Y_{1}+Y_{1}Z_{2}\right)+\left(Y_{0}Z_{1}Z_{2}+Z_{0}Z_{1}Y_{2}\right)
\end{equation}

This dynamics can be Trotterized with 8 two-qubit (CZ) gates per Trotter step and about 10 single-qubit gates (the details are in Appendix D), which means that on IBMQ machines we can preform only 2 or 3 Trotter steps within $10\%$ of the coherence time. Unfortunately this is not enough for adiabatic ground state preparation with acceptable fidelities, so we focus on time evolution with an initial excitation (similar to \figref{fig:1d_time_evolution}), that allows us to observe qualitative features of the original model. In this experiment we begin by exciting qubit 1, which corresponds (in the chosen sector) to an initial state with $E_{1} =N_{1} =1$, and $ E_{0} = E_{2} = N_{0} = N_{2} = N_{3} =0$ (see \figref{fig:2d_layout}). The resulting time evolution (\figref{fig:2d_time_evolution}) is similar to the one-dimensional case in the sense that again we observe  different behavior for different values of $h/J$ given the same initial excitation, whose robustness to the dynamics may be qualitatively related to confinement or deconfinement.

%%%%%%%%%%%%%%%%%%%%%%%%%%%%%%%% CONCLUSION %%%%%%%%%%%%%%%%%%%%%%%%%%%%%%%%%%%
\begin{figure} 
    \includegraphics[width = 1\columnwidth]{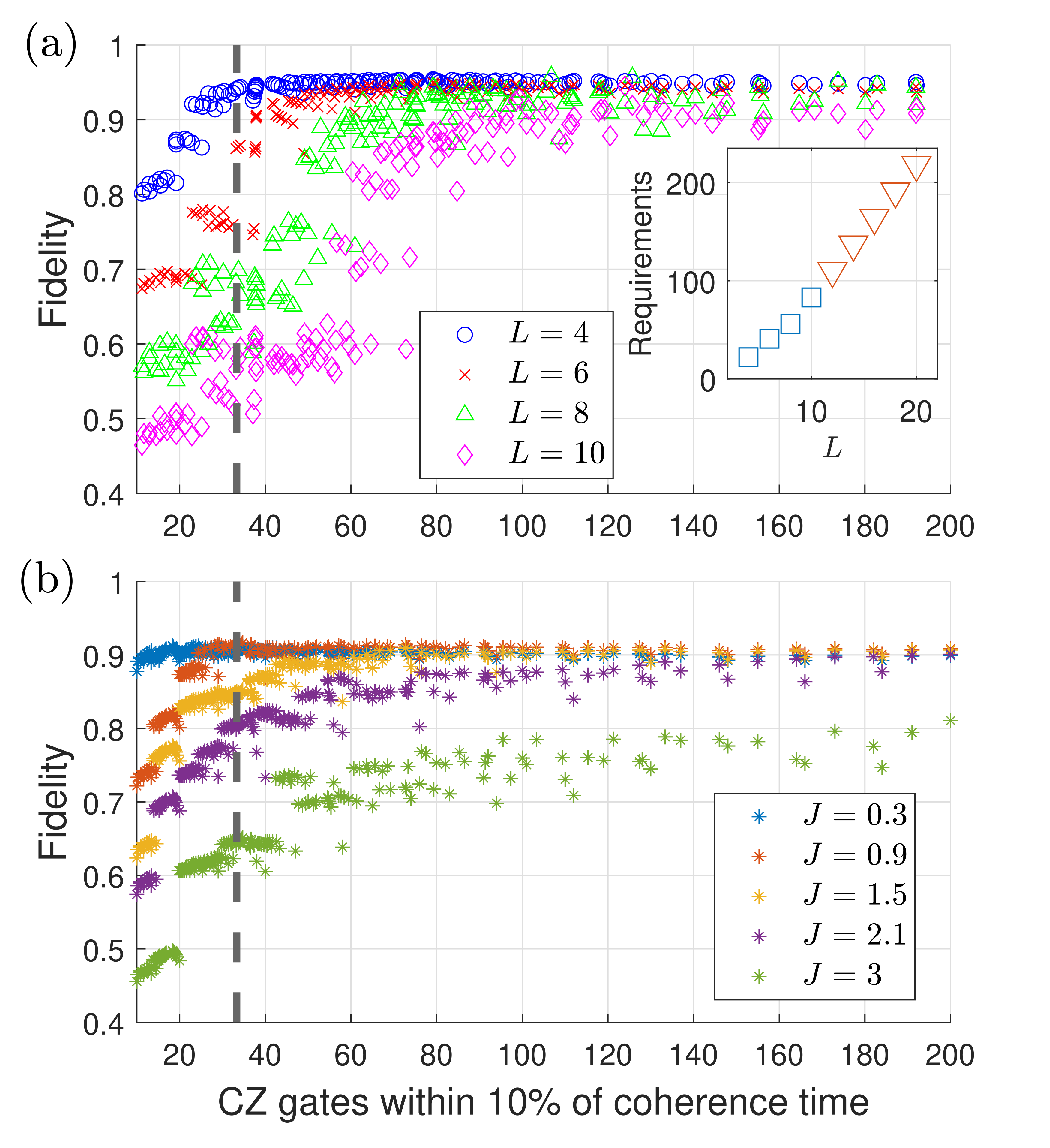}
    \centering
    \caption{Fidelity of the ground-state for the $1+1d$ model obtained via a noisy numerical simulation of the adiabatic ground-state preparation experiment, compared against exact diagonalization. (a) for $J=h=1$ and different values of the system size $L$, and (b) for $L=4$, and different values of $J$ (with $h=1$). Each  data point corresponds to a different value for the coherence time (assuming $T_1=T_2$) and CZ gate duration, and the horizontal axis is the number of CZ gates that can be preformed in $10\%$ of the coherence time (which is a measure of the quality of the quantum hardware). The dashed vertical line represents typical values for IBMQ machines. The inset in (a) shows the hardware requirements, defined as the number of CZ gates in $10\%$ of the coherence time that is required for a ground-state fidelity of at least $90\%$; for different values of $L$. The blue squares are derived from the numerical data of (a), and the red triangles are linear extrapolation.
    }
    \label{fig:numerics}
\end{figure}
\section{Discussion and Summary} \label{discussion_and_summary}
In order to assess the feasibility of our method, we discuss the hardware requirements for more advanced implementations. First, in order to simulate the truly $2+1d$ version of the model, the qubits have to be arranged on a square lattice. As such devices already exist (e.g. the famous Google Sycamore \cite{google_supremacy_2019}), this particular technical problem can be considered solved. Next, we consider what are the requirements on the quality of the qubits for extending beyond the proof-of-concept demonstrations shown in this work. For that we numerically simulate the adiabatic ground-state preparation experiment described above (section \ref{1d_experiment}) - with a trivial initialization of the $J=0$ ground-state and an adiabatic increase of $J$ through the Trotter steps up to some finite value.

We do this using different values for the qubits' coherence time and the CZ gate-duration, and record the fidelity of the final state relative to the exact ground state obtained by diagonalization. Taking $J=h=1$ and different values of $L$, we can conclude (\figref{fig:numerics}(a)) that while going beyond $L=4$ is challenging for the IBMQ devices, it should be possible with slightly longer coherence times or faster gates  that are achievable on other platforms \cite{Quantinuum_2022}. While a physical interpretation as a $\mathbb{Z}_2$ LGT is only valid when $L$ is an integer multiple of $4$ (see section \ref{method2}), evolving the simulation Hamiltonian \eqref{eqn:sim_H_1D} for other values of $L$ is still a valid approach to assessing how well the method scales with the system size. Similarly, \figref{fig:numerics}(b) shows the hardware requirements for probing the $L=4$ model with intermediate values of $J/h$ (recall that it is possible to start the adiabatic ground-state preparation procedure from either the $h=0$ or $J=0$ direction). The jumps in the plots are an artifact of the choice of a different number of Trotter steps for each data point in an attempt to optimize Trotter errors against decoherence errors. In principle, the total simulated time (which determines the adiabaticity of the process) could also be optimized for different values of $J$ and noise parameters. However in this case we work with a single value for the sake of simplicity, which results in the fidelity saturating at a value that is smaller than $1$.

To conclude, in this work we have presented a way to overcome two major bottlenecks in the quantum simulation of LGTs: one being the challenge of simulating fermionic matter and the other is the redundancy of the Hilbert space. We have shown how both of these problems can be tackled by solving the local constraints (Gauss' law) for the matter rather than for the gauge field.

 To compare our method with the more conventional ones, we demonstrated it for the simplest case of $\mathbb{Z}_2$, in a one space dimension. In the conventional methods one uses the non-local Jordan-Wigner map; or solves the local constraints for the gauge field, which introduces non-locality; or both. Our method avoids both these techniques and thus extends easily to higher dimensions without non-locality, which we demonstrated for $d=2$. Furthermore, the method can be applied to more complicated gauge groups, including non-Abelian ones (specifically $U\paren{N}$ and $SU\paren{N}$), following the criteria worked out in Refs. \cite{zohar_eliminating_2018,zohar_removing_2019}, and used as a basis for quantum simulation, depending on the availability of suitable platforms (in terms of dimensionality and the Hilbert spaces required for the different gauge groups). In any case, the constraints and redundancy are eliminated, making the feasibility question technological and not conceptual. 
 
 Our scheme thus imposes fairly modest requirements on the simulator: no redundant components, no local constraints to maintain and no need to directly implement fermions. Moreover, we have shown how to implement it in one dimension using only simple single- and two-body operations, which we demonstrated experimentally on the IBMQ platform. While the platform is not suitable for implementing our method for the fully two-dimensional model, we demonstrated it on a quasi-two dimensional toy-model which is a minimal version of the theory for which traditional methods fail.
 
 Because of these modest requirements we believe that our method could become an important and useful tool for quantum simulation of LGTs with fermionic matter. As quantum technology progresses, we expect that in the near future it will be applied to real unsolved problems rather than mere demonstrations. 

%%%%%%%%%%%%%%%%%%%%%%%%%%%%%%%% ACKNOWLEDGMENTS %%%%%%%%%%%%%%%%%%%%%%%%%%%%%%%%%%%
\section*{Acknowledgments}
We acknowledge the use of IBM Quantum services for this work and to advanced services provided by the IBM Quantum Researchers Program. The views expressed are those of the authors, and do not reflect the official policy or position of IBM or the IBM Quantum team. This research was supported by the Israel Science Foundation
(grant no. 523/20 and 2323/19).

G.P. and T.G. contributed equally to this work.

%%%%%%%%%%%%%%%%%%%%%%%%%%%%%%%% APPENDICES %%%%%%%%%%%%%%%%%%%%%%%%%%%%%%%%%%%
\section*{Appendix A: Applying the matter removal transformation } \label{app_a}
Here, we give further details about the matter removal process, going from the hard-core bosonic setting to the one without matter at all. This should be read as a more detailed description of the process outlined in section \ref{method2}.

It is straightforward to verify that the $\mathcal{U}^{(2)}$ transformation (defined in section \ref{method2})  gives rise to
\begin{equation} \label{eqn:U2_relations}
\begin{aligned}
\mathcal{U}^{(2)}\sigma^{\pm}_{n}\mathcal{U}^{(2)\dagger} &= P^{+}_n \sigma^{\mp}_n + P^{-}_n \sigma^{\pm}_n\\
\mathcal{U}^{(2)}\sigma^{z}_{n}\mathcal{U}^{(2)\dagger} &= \varepsilon_n S_n \sigma^z_n\\
\mathcal{U}^{(2)}X_{n}\mathcal{U}^{(2)\dagger} &= \sigma^x_n X_n \sigma^x_{n+1}\\
\mathcal{U}^{(2)}Z_{n}\mathcal{U}^{(2)\dagger} &=  Z_n 
\end{aligned}
\end{equation}
which results in the transformed constraints of \fulleqref{eqn:gauss_law_method2}
simply by acting with $\mathcal{U}^{(2)}$ on the hard-core bosonic version of Gauss' law (\fulleqref{eqn:gauss_law_method1}).

Acting with $\mathcal{U}^{(2)}$ on the Hamiltonian, we find that the electric part is invariant,
\begin{equation}
	H^{(2)}_{\text{E}}=\mathcal{U}^{(2)}H^{(1)}_{\text{E}}\mathcal{U}^{(2)\dagger}= - h\underset{n}{\sum}Z_n
\end{equation}

Considering the transformation of the mass Hamiltonian, we take a step back, and consider its effective form in the chosen sector. Using the relevant Gauss' law constraints (\ref{eqn:gauss_law_method2}) 
before applying $\mathcal{U}^{(2)}$, the mass Hamiltonian effectively takes the form
\begin{equation}
	H^{(1)}_{\text{m}}=\frac{m}{2}\underset{n}{\sum}\left(-1\right)^{n}\sigma^z_n \underset{\text{eff}}{=}
	-\frac{m}{2} \underset{n}{\sum} \paren{-1}^n \varepsilon_n S_n.
\end{equation}
As $S_n = Z_{n-1}Z_n$, this is invariant under the transformation $\mathcal{U}^{(2)}$, so we have the the effective expression 
\begin{equation}
	H^{(2)}_{\text{m}}=\mathcal{U}^{(2)}H^{(1)}_{\text{m}}\mathcal{U}^{(2)\dagger}\underset{\text{eff}}{=}-\frac{m}{2} \underset{n}{\sum} \paren{-1}^n \varepsilon_n Z_{n}Z_{n+1}
\end{equation}
for the transformed mass Hamiltonian (and "effective" here means "in the chosen sector").

Finally, consider the transformation of the interaction part. Here, we have terms like $Z_{n-1} \sigma_n^+ X_n \sigma_{n+1}^-$ (and its Hermitian conjugate,  see \fulleqref{eqn:H_1_1d}) that have to be transformed under $\mathcal{U}^{(2)}$, resulting in (using \fulleqref{eqn:U2_relations})
\begin{equation}
         Z_{n-1}
 	\left(P^{+}_n \sigma^{-}_n + P^{-}_n \sigma^{+}_n\right)
 	\sigma^x_n X_n \sigma^x_{n+1}
 \left(P^{+}_{n+1} \sigma^{+}_{n+1} + P^{-}_{n+1} \sigma^{-}_{n+1}\right).
\end{equation}
Noting that 
$\sigma^{\pm}\sigma^x = \frac{1}{2}\left(1\pm\sigma^{z}\right)$ is a projection operator to $\sigma^z=\pm1$ and that
$\sigma^x\sigma^{\pm} = \frac{1}{2}\left(1\mp\sigma^{z}\right)$ is a projection operator to $\sigma^z=\mp1$, we see that the transformed Hamiltonian is block-diagonal in the matter spins, with static $\sigma_z$ configurations. Having the constraints (\ref{eqn:gauss_law_method2}) in mind, we can restrict ourselves to physical states by ignoring all the terms but those that are projectors onto matter down-states ($\sigma^z=-1$), and then neglect the matter spins altogether. Formally, \fulleqref{eqn:gauss_law_method2} implies that
\begin{equation}
	\left|\psi^{(2)}\right\rangle = \left|\tilde{\psi}^{(2)}\right\rangle \otimes \left|\text{out}\right\rangle
\end{equation}
where $\left|\text{out}\right\rangle$ is a product state of all the matter spins in which they all point down, and $\left|\tilde{\psi}^{(2)}\right\rangle$ is a state of the gauge fields. Then, we can define a Hamiltonian acting only on the gauge fields degrees of freedom, by
\begin{equation}
	\tilde{H}^{(2)}= \left\langle \text{out} \right| H^{(2)}\left|\text{out}\right\rangle
\end{equation}
including $\tilde{H}^{(2)}_{\text{m}} = H^{(2)}_{\text{m}}$, 
$\tilde{H}^{(2)}_{\text{E}} = H^{(2)}_{\text{E}}$ and
\begin{equation}
\begin{split}
\tilde{H}^{(2)}_{\text{GM}}&= 
iJ\underset{n}{\sum}Z_{n-1}
P^{+}_n X_n 
 P^{+}_{n+1}  +\text{h.c.}\\
& =
 i\frac{J}{4}\underset{n}{\sum}Z_{n-1}
 \left[X_n + \varepsilon_n\left(S_nX_n - X_nS_{n+1}\right)-S_nX_nS_{n+1}\right] +\text{h.c.}
 \end{split}
\end{equation}
If we assume that $J$ is real (for $d=1$ this can be assumed without loss of generality), the expression will be simplified as only anti-Hermitian contributions in the sum would have to be considered. This eventually leads to the expression \eqref{eqn:H_GM_tilde_2} in section \ref{method2}.

\section*{Appendix B: Trotter-error analysis} \label{app_b}
We would like to approximate the time evolution under a Hamiltonian broken to three pieces,
\begin{equation}
H=H_1+H_2+H_3
\end{equation}
by the Trotterized sequence \cite{trotter_on_1959,suzuki_decomposition_1985}
\begin{equation}
e^{-iHt} \approx \left(e^{-i\epsilon H_1}e^{-i\epsilon H_2}e^{-i\epsilon H_3}\right)^{\mathcal{N}},
\end{equation}
where $\mathcal{N}=t/\epsilon$ is a very large integer. The choice of $\mathcal{N}$, as usual, is a compromise between the experimental capabilities and the Trotterization error that we allow.

Suppose we allow some error $\delta$. Then, we would like to have
\begin{equation}
\|e^{-iHt} - \left(e^{-i\epsilon H_1}e^{-i\epsilon H_2}e^{-i\epsilon H_3}\right)^{\mathcal{N}}\|\sim\delta
\end{equation}
which can be bounded, in leading order, by \cite{suzuki_decomposition_1985} 
\begin{equation}
\delta \lesssim \frac{t^2}{2\mathcal{N}}\|\underset{i<j}{\sum}\left[H_i,H_j\right]\|.
\end{equation}

In our case, we have
\begin{equation}
	\begin{aligned}
	\left[\hat{H}_{\text{GM}},\hat{H}_{\text{m}}\right]&=
	\frac{imJ}{2}\underset{n}{\sum}X_n\left(Z_{n-1}+Z_n\right)\\
	\left[\hat{H}_{\text{GM}},\hat{H}_{\text{E}}\right]&=
	ihJ\underset{n}{\sum}Z_{n-1}X_nZ_{n+1}
	\\&-\frac{iJ^2}{2}\underset{n}{\sum}X_n\left(Z_{n-2}Y_{n-1}+Y_{n+1}Z_{m+2}\right)\\
	\left[\hat{H}_{\text{m}},\hat{H}_{\text{E}}\right]&=-\frac{imJ}{2}\underset{n}{\sum}X_n\left(Z_{n-1}+Z_n\right).
	\end{aligned}
\end{equation}
Clearly, the first and third commutator cancel each other, and thus the order choice presented in section \ref{trotterization_etc},
\begin{equation}
	e^{-i\hat{H}t}\approx \left(e^{-i\epsilon H_{\text{GM}}}e^{-i\epsilon H_{\text{m}}}e^{-i\epsilon H_{\text{E}}}\right)^{\mathcal{N}}=\left(\Omega_{\text{GM}}\Omega_{\text{m}}\Omega_{\text{E}}\right)^{\mathcal{N}},
\end{equation}
would be optimal.

Upon computing the norms, we get that the error is of order
\begin{equation}
	\delta \sim \frac{t^2}{2\mathcal{N}}\left(J^2 + \left|Jh\right|\right)L,
\end{equation}
implying that a reasonable $\mathcal{N}$ to use, given $\delta$ and the parameters $h,j$ (independently of $m$) is
\begin{equation}
	\mathcal{N} \sim \frac{t^2}{2\delta}\left(J^2 + \left|Jh\right|\right)L
\end{equation}

\section*{Appendix C: Measuring non-local observables} \label{app_c}
A valid question in the design of a quantum simulator is what are the observables one is interested in measuring, and how to measure them. In LGTs, the relevant quantities are the gauge invariant observables. In section \ref{trotterization_etc} we discuss the local ones, namely the number operator and the electric field operator, and explain how to measure them within our quantum simulation scheme. Here we focus on the the non-local ones, namely the \emph{mesonic string operators},
\begin{equation}
	\mathcal{M}\left(n,n+R\right)=\psi^{\dagger}_n \left[\overset{n+R-1}{\underset{m=n}{\prod}}X_m\right] \psi_{n+R},
\end{equation}
for $R\ge 1$ ($R=0$ gives the local number operator).
There are two questions to be asked; first, what are transformed operators  $\hat{\mathcal{M}}\left(n,n+R\right)$, whose measurement in the simulated physical states $\left|\hat{\psi}\right\rangle$ will correspond to measuring the original operators with respect to the original states $\left|\psi\right\rangle$? In other words, what are the  $\hat{\mathcal{M}}\left(n,n+R\right)$ for which
\begin{equation}
% \left\langle \hat{\psi}\right| \hat{Z}_n \left|\hat{\psi}\right\rangle&=\left\langle \psi\right| Z_n \left|\psi\right\rangle\\
\left\langle \hat{\psi}\right| \hat{\mathcal{M}}\left(n,n+R\right) \left|\hat{\psi}\right\rangle =\left\langle \psi\right| \mathcal{M}\left(n,n+R\right) \left|\psi\right\rangle.
\end{equation}
The second question would be how to actually perform such measurements in our simulator.

To answer the first question, reformulating $\mathcal{M}\left(n,n+R\right)$ using hard-core bosons by applying $\mathcal{U}^{(1)}$ results in \cite{zohar_eliminating_2018}: 
\begin{equation}
	\mathcal{M}^{(1)}\left(n,n+R\right)=i\left(-1\right)^R Z_{n-1} \sigma^{+}_n 
	\left[\overset{n+R-2}{\underset{m=n}{\prod}}Y_m\right]X_{n+R-1} \sigma^{-}_{n+R} 
\end{equation}
(when $R=1$, the product of $Y_m$ is not included).
Acting on this with $\mathcal{U}^{2}$, projecting onto $\left|\text{out}\right\rangle$ and 
rotating with $\mathcal{V}$, one finds the relevant observable to measure (given here in the sector $\varepsilon_n=\paren{-1}^n$):

\begin{equation}
\begin{split}
	\hat{\mathcal{M}}\paren{n,n+R} &= \mathcal{V} \left\langle \text{out}\right| \mathcal{U}^{(2)} 
	\mathcal{M}^{(1)}\left(n,n+R\right) 
	\mathcal{U}^{(2)\dagger}
	\left|\text{out}\right\rangle
	\mathcal{V}\\ 
	&\equiv \frac{1}{4}\left(-1\right)^{R\left(2n+R-1\right)/2}\overset{4}{\underset{\alpha=1}{\sum}}
	\hat{\mathcal{M}}_{\alpha}\left(n,n+R\right). 
\end{split}
\end{equation}

The factor $\left(-1\right)^{R\left(2n+R-1\right)/2}$ is irrelevant; and the mesonic string expectation value will be obtained from measuring the expectation value of the four terms $\hat{\mathcal{M}}_{\alpha}\left(n,n+R\right) $,
\begin{equation}
	\begin{aligned}
\hat{\mathcal{M}}_{1}\left(n,n+R\right) &= Z_{n-1}\left[\overset{n+R-2}{\underset{m=n}{\prod}}Y_m\right]X_{n+R-1}  \\
\hat{\mathcal{M}}_{2}\left(n,n+R\right) &= i\left(-1\right)^n X_{n}\left[\overset{n+R-2}{\underset{m=n+1}{\prod}}Y_m\right]X_{n+R-1}  \\
\hat{\mathcal{M}}_{3}\left(n,n+R\right) &= i\left(-1\right)^{n+R} Z_{n-1}\left[\overset{n+R-1}{\underset{m=n}{\prod}}Y_m\right]Z_{n+R} \\
\hat{\mathcal{M}}_{4}\left(n,n+R\right) &= \left(-1\right)^{R+1} X_{n}\left[\overset{n+R-1}{\underset{m=n+1}{\prod}}Y_m\right]Z_{n+R}
	\end{aligned}
\end{equation}
They are all products of $X,Y,Z$ operators along the string, with spectrum $\pm 1$, and they can be measured, for example, by using an ancillary qubit interacting sequentially along the string with all the links, one after the other, and using the right controlled gates accumulating their $X$, $Y$ or $Z$ contribution to the product (this is not a new method - see, e.g., \cite{zohar_local_2020} for details).

\section*{Appendix D: Trotterization of the quasi two-dimensional Hamiltonian}  \label{app_d}
We have to construct a Trotterization  of \fulleqref{eqn:sim_H_2D} based on single-qubit rotations and two-qubit gates between neighbouring qubits on a three qubits chain. The non trivial terms are the three-qubit interactions $Y_0 Z_1 Z_2$, $Z_0 Z_1 Y_1$ and the two-qubit terms $Z_0 Y_1$, $Y_1 Z_2$. Naively, we can implement each of these four terms using \fulleqref{eqn:CZ_property} and the analogous properties of the controlled-NOT (CX) gate, resulting in:

\begin{align}
    Y_0 Z_1 Z_2 &= \text{CX}^{21} \text{CZ}^{10} Y_0 \text{CZ}^{10} \text{CX}^{21} \label{eqn:2Dtrtter_1}\\ 
    Z_0 Z_1 Y_2 &= \text{CX}^{01} \text{CZ}^{12} Y_2 \text{CZ}^{12} \text{CX}^{21}\label{eqn:2Dtrtter_2}\\
     Y_1 Z_2 &=  \text{CZ}^{12} Y_1 \text{CZ}^{12} \label{eqn:2Dtrtter_3}\\
     Z_0 Y_1 &=  \text{CZ}^{01} Y_1 \text{CZ}^{01},\label{eqn:2Dtrtter_4} 
\end{align}
where $\text{CX}^{mn}$ and $\text{CZ}^{mn}$ are controlled-NOT and controlled-Z operators between qubits $m$ and $n$. This implementation requires $12$ two-qubits gates per trotter step, so it is somewhat inefficient and we can do better. 
Instead of treating each of the four non-trivial terms separately, we note that we can implement two of them together, as, for example, one can verify that the transformation $\text{CZ}^{12} \text{CX}^{21} \text{CZ}^{10}$ takes $Y_0$ to $Y_0 Z_1 Z_2$ and $Y_1$ to $Z_0 Y_1$.
This means that we can implement the terms $Y_0 Z_1 Z_2$ and  $Z_0 Y_1$ together, by transforming with $\text{CZ}^{12} \text{CX}^{21} \text{CZ}^{10}$ and rotating qubits $1$ and $0$ around the $Y$ axis: 
\begin{equation} 
    e^{i\frac{J}{2}\paren{Y_0 Z_1 Z_2 + Z_0 Y_1}t}\approx \paren{\text{CZ}^{12}\text{CX}^{21} \text{CZ}^{10} U^0_Y\paren{\epsilon} U^1_Y\paren{\epsilon} \text{CZ}^{10} \text{CX}^{21} \text{CZ}^{12}}^\mathcal{N}, 
\end{equation}
where $U^n_Y\paren{\epsilon} = \exp{\paren{i\epsilon J Y_n /2}}$, and a similar result holds for the other two terms: $Z_0 Z_1 Y_2 + Y_1 Z_2$.
These two steps together still use $12$ two-qubit gates, but in this form the trotter step can be simplified further, since a combination of two entangling (two-qubit) gates on the same pair of qubits can be decomposed into a single entangling gate and a few single-qubit rotations. Using this kind of decompositions one can implement our trotter step with only $8$ two-qubit gates, and 10 single qubit steps. The single-qubit depth can be larger if a general single-qubit rotation is not a native gate on the platform (as is the case for IBMQ devices), but on the other hand it is reasonable to assume that it can also be reduced with circuit optimization. We do not think it is useful to discuss this further here since two-qubit gates are the major source of error in current devices.

\bibliography{ref}

\end{document}